\documentclass[useAMS,usenatbib]{mn2e}

\usepackage{amssymb}
\usepackage{amsmath}
\usepackage{txfonts}
\usepackage{graphicx}
\usepackage{natbib}
\usepackage{rotating}
\usepackage{url}
\usepackage{epsfig}
\usepackage{xspace}
 
\def\bi {\begin{itemize}}
\def\ei {\end{itemize}}

\def\xmm{\textit {XMM-Newton}\xspace}

\def\deg{$^\circ$}
\def\lsi {LSI~+61\deg~303\xspace}
\def\szk{\textit {Suzaku}}

\def\sw{\textit{Swift}\xspace}
\def\nh{$N_H$\xspace}

\usepackage{color}
\definecolor{red}{rgb}{0.7,0,0}
\definecolor{blue}{rgb}{0,0,0.7}

\begin{document}
%\end{document}

\title{Study of  orbital and superorbital variability of \lsi\ with X-ray data} 

 \author[M. Chernyakova et al.]{M. Chernyakova$^{1,2}$, Iu. Babyk$^{1,2,3,8}$,  D. Malyshev$^{4}$, Ie. Vovk$^{5}$, S. Tsygankov$^{6}$,  \newauthor
  H. Takahashi$^{7}$, Ya. Fukazawa$^{7}$\\
$^1$ School of Physical Sciences, Dublin City University, Dublin 9, Ireland\\
$^2$ Dublin Institute for Advanced Studies, 31 Fitzwilliam Place, Dublin 2, Ireland\\
$^3$ University of Waterloo, 200 University Ave W, Waterloo, Ontario, N2L 3G1, Canada\\
$^4$ Institut f{\"u}r Astronomie und Astrophysik T{\"u}bingen, Universit{\"a}t T{\"u}bingen, Sand 1, D-72076 T{\"u}bingen, Germany\\
$^5$  Max Planck Institute for Physics, F{\"o}hringer Ring 6, 80805 Munich, Germany\\
$^6$ Tuorla Observatory, Department of Physics and Astronomy,  University of Turku,  V\"ais\"al\"antie 20, FI-21500, Piikki\"o, Finland\\
$^7$ Department of Physical Science, Hiroshima University, 1-3-1 Kagamiyama, Higashi-Hiroshima, Hiroshima, 739-8526, Japan\\
$^8$ Main Astronomical Observatory of National Academy of Science of Ukraine, Academica Zabolotnogo str., 27, Kyiv, 03143, Ukraine 
}
\date{Received $<$date$>$  ; in original form  $<$date$>$ }
\pagerange{\pageref{firstpage}--\pageref{lastpage}} \pubyear{2015}
\maketitle
\label{firstpage}

\begin{abstract} \lsi is one of the  few X-ray binaries with a Be star
companion  from which  radio, X-rays  and high-energy gamma-ray (GeV and TeV) emission 
have been observed.  The nature of the high energy activity of the system is not yet fully understood, but it is widely believed that it is generated due to the interaction of the relativistic electrons leaving the compact object with the photons and non-relativistic wind of the Be star.    The superorbital variability  of the system has been observed in the radio, optical and X-ray domains and could be due to the cyclic change of the Be star disk size. In this paper we systematically review all publicly available data from \textit{Suzaku, XMM-Newton, Chandra} and  \textit{Swift} observatories in order to measure the absorption profile of the circumstellar Be disk as a function of orbital and superorbital phases. We also discuss short-term variability of the system, found during the analysis and its implications for the understanding of the physical processes in this system.
\end{abstract}


%%%%%%%%%%%%%%%%%%%%%%%%%%%%%%%%%%%%%%%%%%%%%%%%%%%%%%%%%%%%%%%%%%%%%%%%%%%%%
\section{Introduction}
%%%%%%%%%%%%%%%%%%%%%%%%%%%%%%%%%%%%%%%%%%%%%%%%%%%%%%%%%%%%%%%%%%%%%%%%%%%%%

\lsi is one of the few X-ray binaries observed from the radio to very high-energy gamma-ray bands. It consists of a Be star and a compact object on an eccentric orbit (see Fig.~\ref{ellips}). The nature of the compact object is not known due to poor constraints on its mass implied by large uncertainty of the inclination angle \citep[see, e.g.,][]{bartosik09,caliandro12}.

The radio observations of \lsi cover more than two decades and demonstrate that the observed emission is modulated on timescales of $\sim 26.5$~d and $\sim 1667$~d~\citep{gregory02}. These timescales are usually interpreted as the orbital and superorbital periods of the system (see, however, \citet{massi13,massi15}). The detection of similar periods has been recently reported in the optical~\citep{zamanov13,fortuny15}, X-ray~\citep{chernyakova12,li14} and gamma-ray bands~\citep{lsifermi13,jaron14}.

The nature of the superorbital period is not well established. \citet{chernyakova12} found that while both, X-ray and radio, lightcurves of \lsi demonstrate one peak per orbit (with the position of orbital phase modulated on the superorbital scale), the X-ray peak systematically outruns the radio one by  about $\sim 5.3$ days. The observed behaviour was interpreted in terms of separate radio and X-ray emission regions. In the dense region of the Be star's equatorial disk, the interaction of the compact object with the wind leads to the injection of high-energy particles into the wind. These particles escape from the system with the stellar wind velocity. While the  X-ray emission can be observed even if it is generated within the system, the radio emission from particles located closer than  $\sim 5\times 10^{13}$~cm to the Be star, is heavily absorbed \citep{zdziarsky10}. The observed phase lag between X-ray and radio maxima corresponds to the escape time of the high-energy particles from the system with the stellar wind. The superorbital period in this model is connected with the processes of gradual build-up and decay of the Be star disk. Shortly after the superorbital phase $\Phi~\sim 0.6$, the disk is weak and can be easily perturbed  and partially striped away by a relativistic outflow from a compact object resulting in a narrow, pronounced X-ray peak. The gradual build-up of the disk up to the superorbital phase $\Phi\sim 0.6$ leads to an increase of the time that the compact object spend in dense regions of the disk and thus to the broadening and blurring of the observed X-ray maximum.

The optical (continuum and $H\alpha$ line) observations of \lsi by \citet{zamanov13,fortuny15} have also revealed the variability of the system on orbital and superorbital timescales. It was found, that the orbital position of the maximum equivalent width of the $H\alpha$ emission line, and the radius of the disk, peak after periastron and coincide on average with the X-ray and $\gamma$-ray maxima. 
%However, the obtained by~\citet{zamanov13} disk's radius does not exceed $\sim 0.5$ periastron distance. This is in the contradiction with \citet{chernyakova12} model, in which at least at some superorbital phases, the compact object has to spend significant time in dense disk regions. \textbf{CHECK THIS TWICE!!??}

Another bit of information regarding  the disk size/density profile can be obtained from low-energy photoelectric absorption in X-rays. Indeed, if the superorbital variability in the system is linked  with the disk build-up process (e.g. \citet{chernyakova12} ) one can expect the gradual increase of the absorption, as the compact object moves on its orbit. 

Below we present an attempt to measure the absorption profile of the circumstellar Be disk as a function of orbital and superorbital phases, based on all available X-ray observations of \lsi. We also discuss short-term variability of the system, found during the analysis and its implications for the understanding of the physical processes in this system.

\section{Data Analysis}
Hereafter, we adopt the following parameters for the \lsi system -- the orbital and superorbital periods $P_{orb} = 26.496 \pm 0.0028$~d and $P_{so}$ = 1667~d are from~\citet{gregory02}. The values for the eccentricity of $e = 0.537\pm0.034$ and the phase of the periastron of $\varphi = 0.275 \pm 0.010$ are adopted from~\citet{aragona09}. Historically the phase of $\varphi = 0$ corresponds to  Julian Date (JD) 2,443,366.775 \citep{gregory02}.

\begin{figure}
\includegraphics[width=1.\columnwidth]{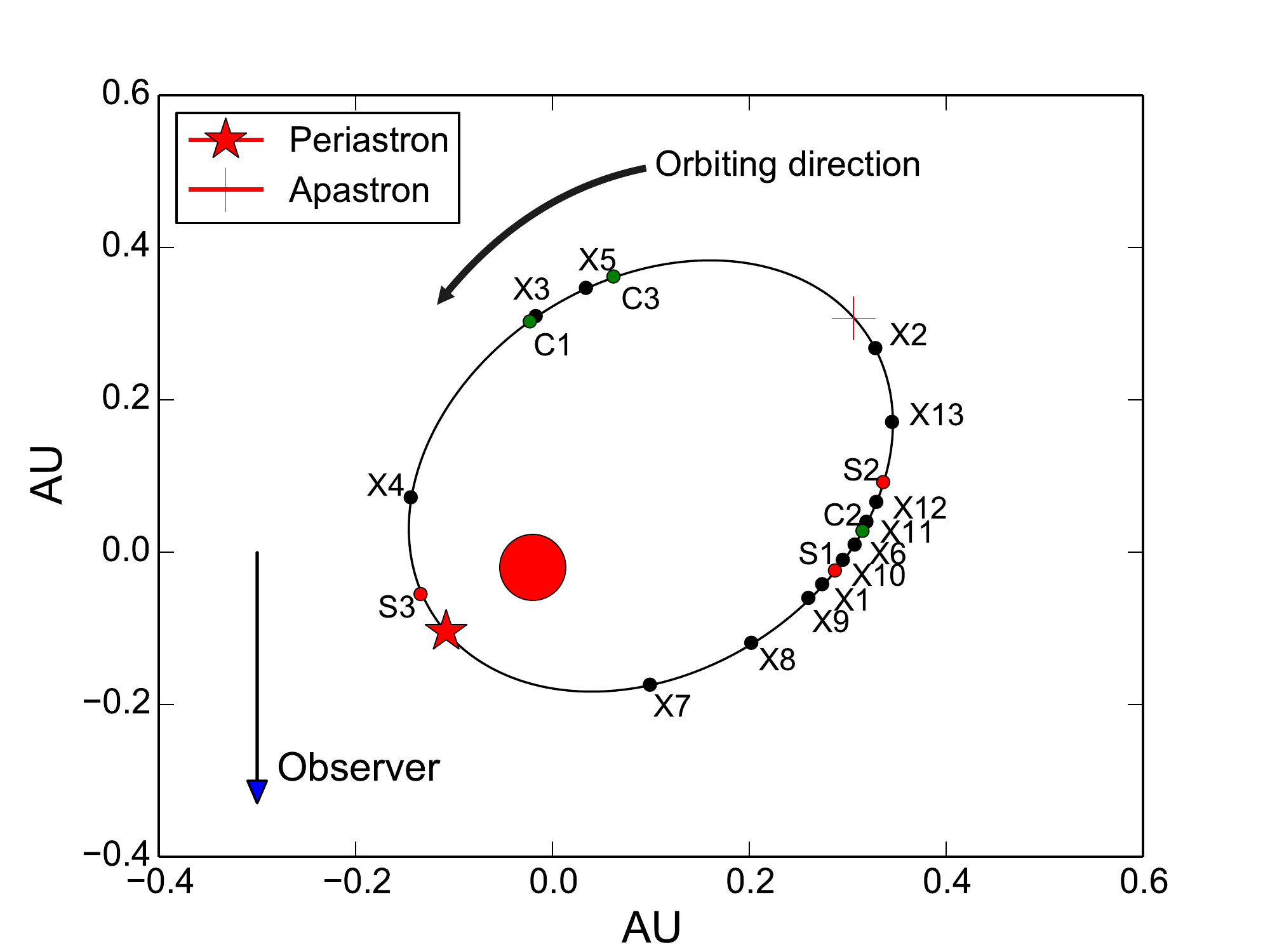}
\caption{Orbital geometry of \lsi\ adapted from Aragona et al. (2009), along with the positions during \szk\ (S), \xmm\ (X), and \textit{Chandra} (C) observations.}
\label{ellips}
\end{figure}

In this work we use all publicly available \szk, \xmm, \textit{Chandra} and \sw data. Graphical representation of the orbital position of all the observations is given in Fig.~\ref{ellips}.
To quantitatively characterise the  short time scale  variability observed at different observations we have calculated the root-mean-square function (RMS) for each light curve. To calculate it we have used a standard definition of  
\begin{equation}
RMS = \sqrt{\frac{\sum(f_{i}-<f>)^{2}}{N}},
\end{equation}
 where $N$ is the full number of time bins in the light curve, and $<f>=\Sigma( f_i \delta f_i^{-2} ) / \Sigma( \delta f_i^{-2} )$ is the weighted mean of the flux with the error of $\delta f_i$ for each time bin $t_i$ with a flux $f_i$.  For consistency we used throughout the paper a time bin size of 1 ks in all the observations, and the corresponded RMS is denoted as RMS$_{1ks}$.

\subsection{\szk\ observations}
In 2009 \szk\ performed three long ($\sim$~170~ks in total) observations of   \lsi\  (see Table~\ref{tab_suz_log}). All these observations occurred within the same orbital cycle. One (S3) happens to be near the periastron and the other two are close to apastron. 

The \szk\ observations were performed with the X-ray Imaging Spectrometer (XIS;  \cite{suzaku_xis}) in the spectral range 0.3 -- 12 keV and the hard X-ray detector (HXD; \cite{suzaku_hxd}) at 13 -- 600 keV. The X-ray imaging spectrometers are located at the focal plane of the XRT and consist of one back-illuminated CCD camera (XIS-1) and three front-illuminated CCDs (XIS-0, -2, and -3). After a fatal damage on 2006 November 9 XIS-2 becomes unusable.
%We do not show HXD-GSO analysis since the GCO do not detect significant signal from \lsi.
\setlength{\tabcolsep}{4pt}
\begin{table}
\caption{Log of \textit{Suzaku}  data. $\phi$ and $\Phi$ are orbital and superorbital  phases correspondingly.}\label{tab_suz_log}
\begin{tabular}{ccccccc}
\hline
Obs. date & ObsID & Ref & MJD & Exp. time & $\phi$ & $\Phi$ \\
 & & &  & ks & &\\
\hline
22-01-2009 & 403015010 & S1 & 54853.951 & 40.5 & 0.562 & 0.8912 \\
25-01-2009 & 403016010 & S2 & 54856.696 & 61.0 & 0.666 & 0.8928 \\
10-02-2009 & 403017010 & S3 & 54872.184 & 68.6 & 0.251 & 0.9021 \\
\hline
\end{tabular}
\end{table}

Data reduction was done using HEASoft\footnote{http://heasarc.gsfc.nasa.gov/docs/software/lheasoft/} v.6.16 software package, while spectral modelling was performed in XSpec environment v.12.8.2. 

For XIS data, we performed full XIS reprocessing and screening data using the \texttt{AEPIPELINE} tool. The processing pipeline V2.1.6.16 was used to process data sets. In our analysis we  used  cleaned event files, with standard screening applied. The source photons were extracted from a circular area with a 3 arc minute radius. The background photons were  taken from a circle with the same radius from a nearby region. The response matrix files (RMF) and auxiliary response files (ARF) were generated using  \texttt{XISRMFGEN} and  \texttt{XISSIMARFGEN}, respectively.

\begin{figure*}
\begin{minipage}{0.33\linewidth}
\includegraphics[width=1.\columnwidth]{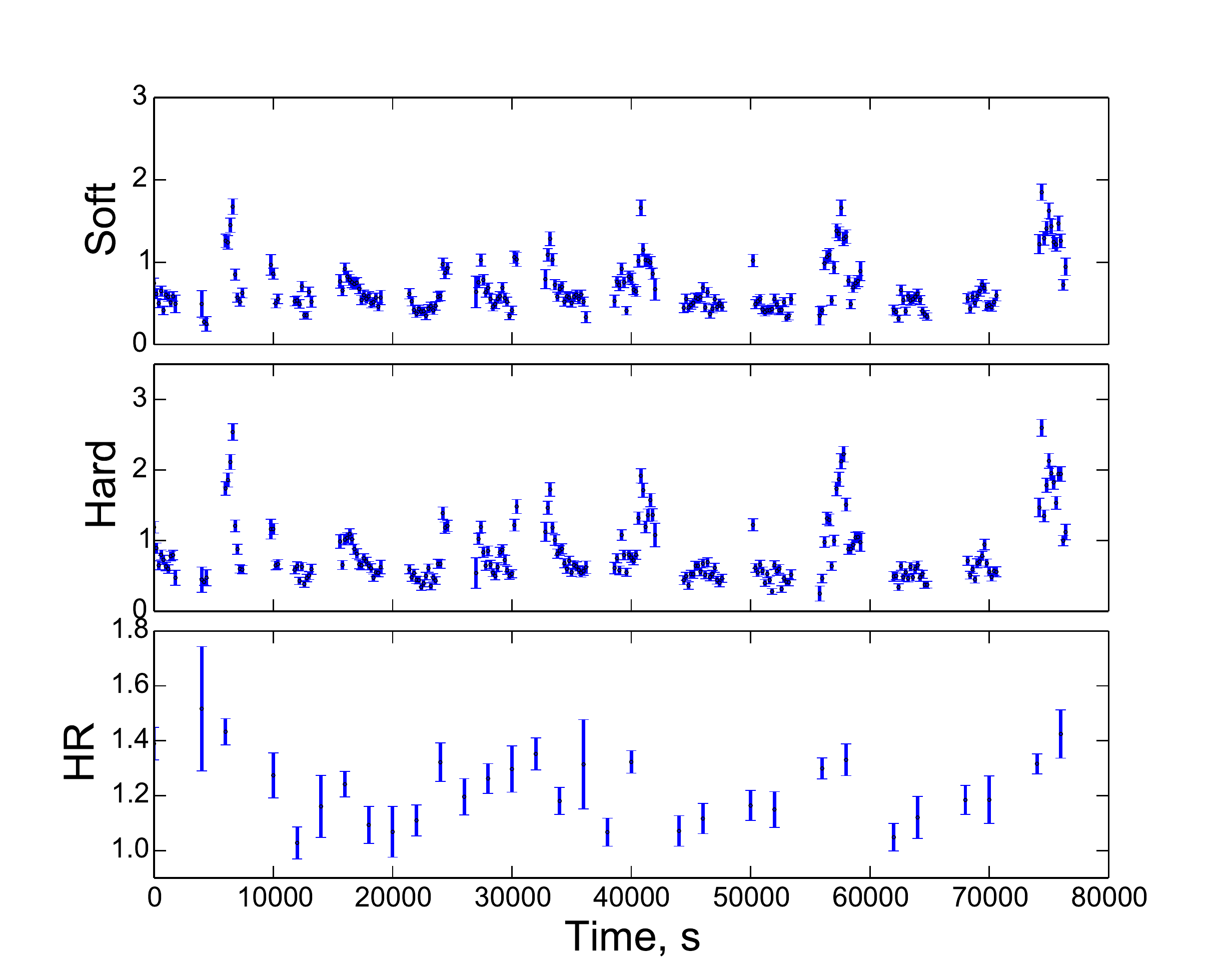}
\end{minipage}
\begin{minipage}{0.31\linewidth}
\includegraphics[width=1.\linewidth]{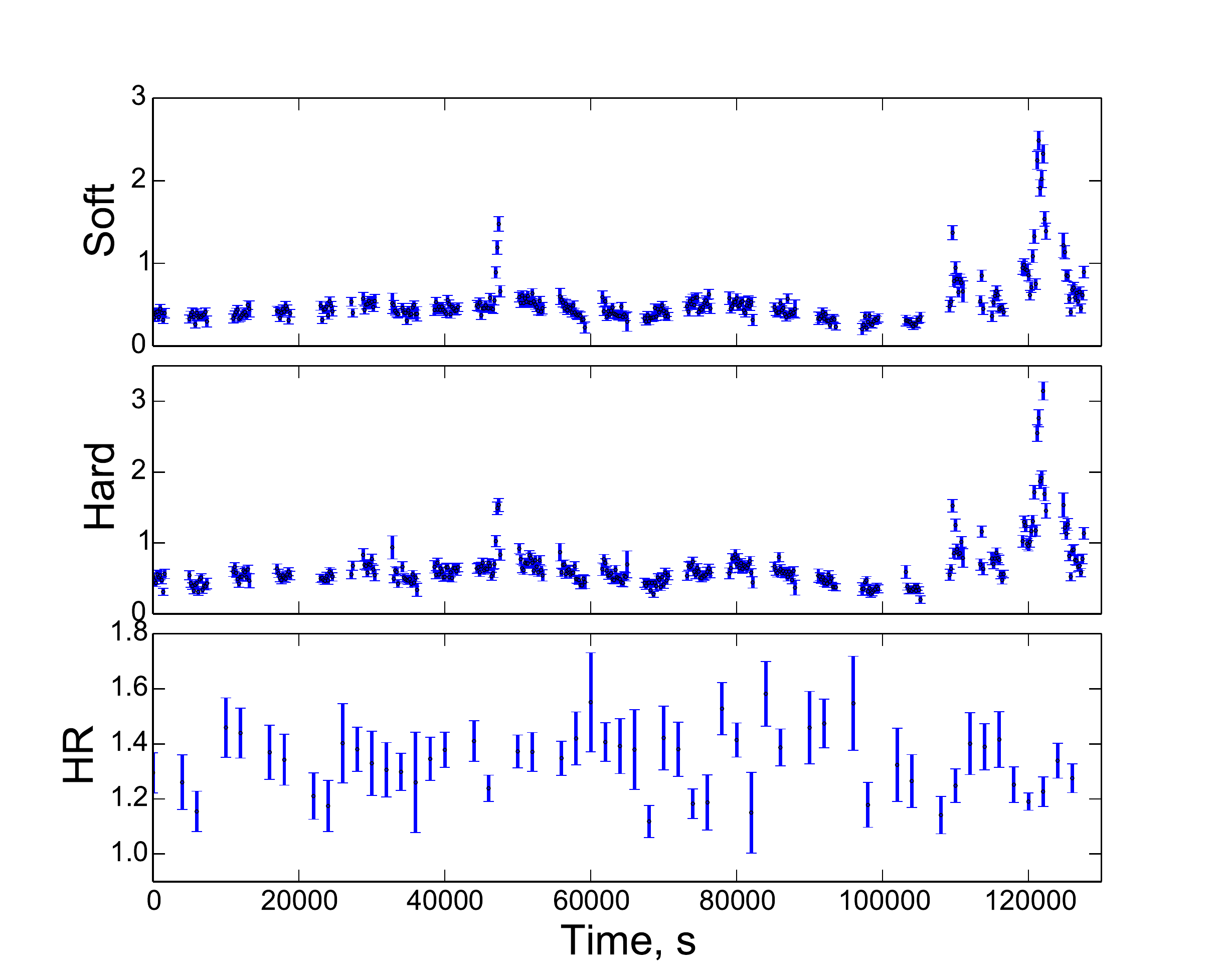}
\end{minipage}
\begin{minipage}{0.33\linewidth}
\includegraphics[width=1.\linewidth]{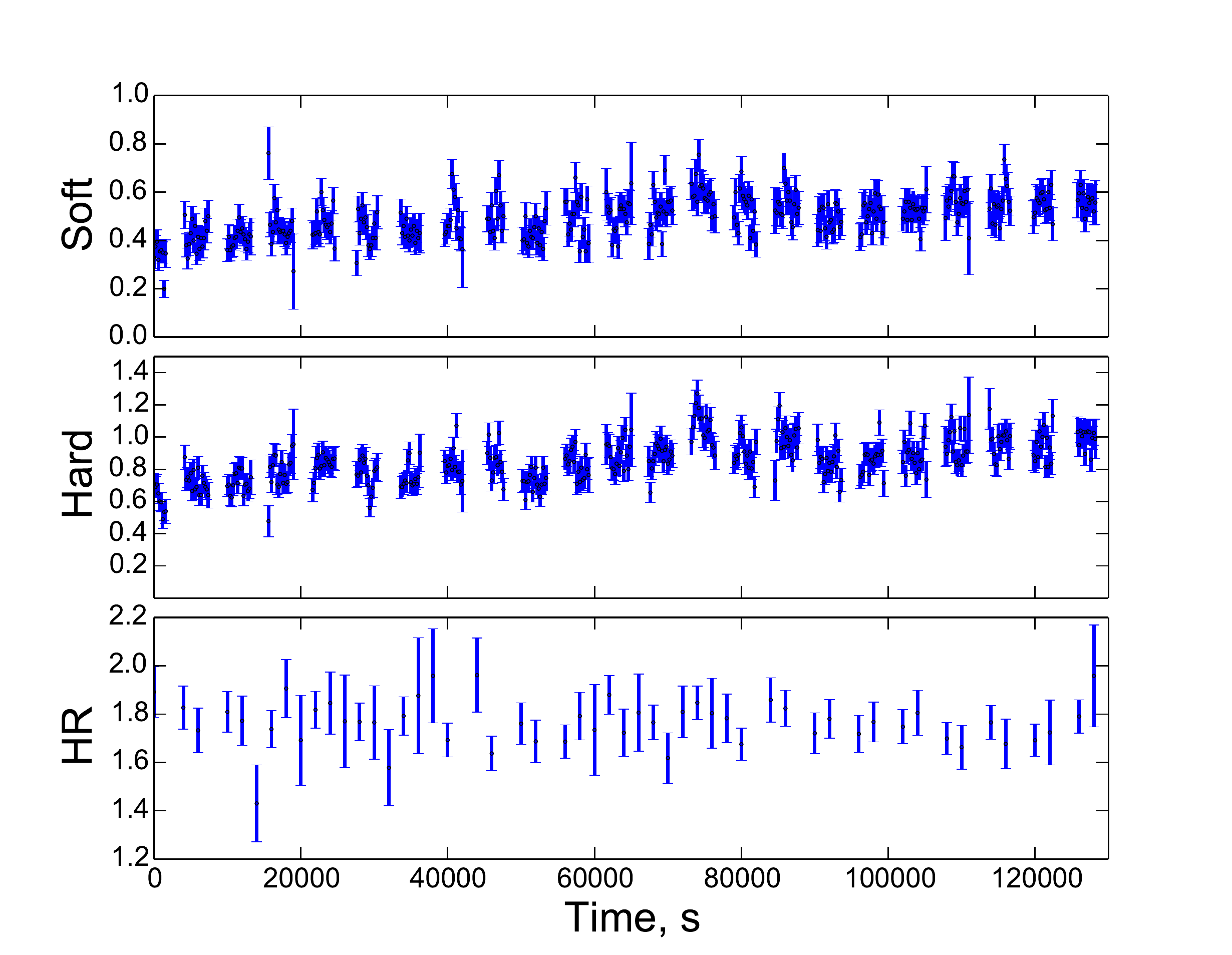}
\end{minipage}
\caption{The Suzaku lightcurves of \lsi\ in two energy bands (soft = 0.3-2.0 keV, hard = 2.0-10.0 keV) together with their hardness ratio (HR = hard/soft) for all three observations (from left to right, S1, S2, S3). Bin time is 200 s for soft and hard ranges and 2 ks for HR.}
\label{fig_lc_suz}
\end{figure*}

Background subtracted light curves obtained from a combination of XIS0, XIS1 and XIS3 detectors are shown in Fig.~\ref{fig_lc_suz} for all three \textit{Suzaku}  observations along with their hardness ratio (HR). Note that while for the lightcurves we have selected a 200 s time bin to resolve flares, HR curves have 2ks time binning chosen to reduce the error.

During both the S1 and S2 observations \szk\ observed a number of short flares during which the flux raised up and subsequently dropped down  by a factor of 5 on a ks time scale.
The S3 observation, done close to the periastron passage,  shows a gradual flux rise, with no flares and a constant hardness ratio during the whole observation. 
{The RMS for each light curve  is given in Table \ref{tab_suz_fit}. It is clearly seen that the S3 observation is much smoother  than S1 and S2 observations. }

\begin{figure}
\centering
\includegraphics[width=\columnwidth]{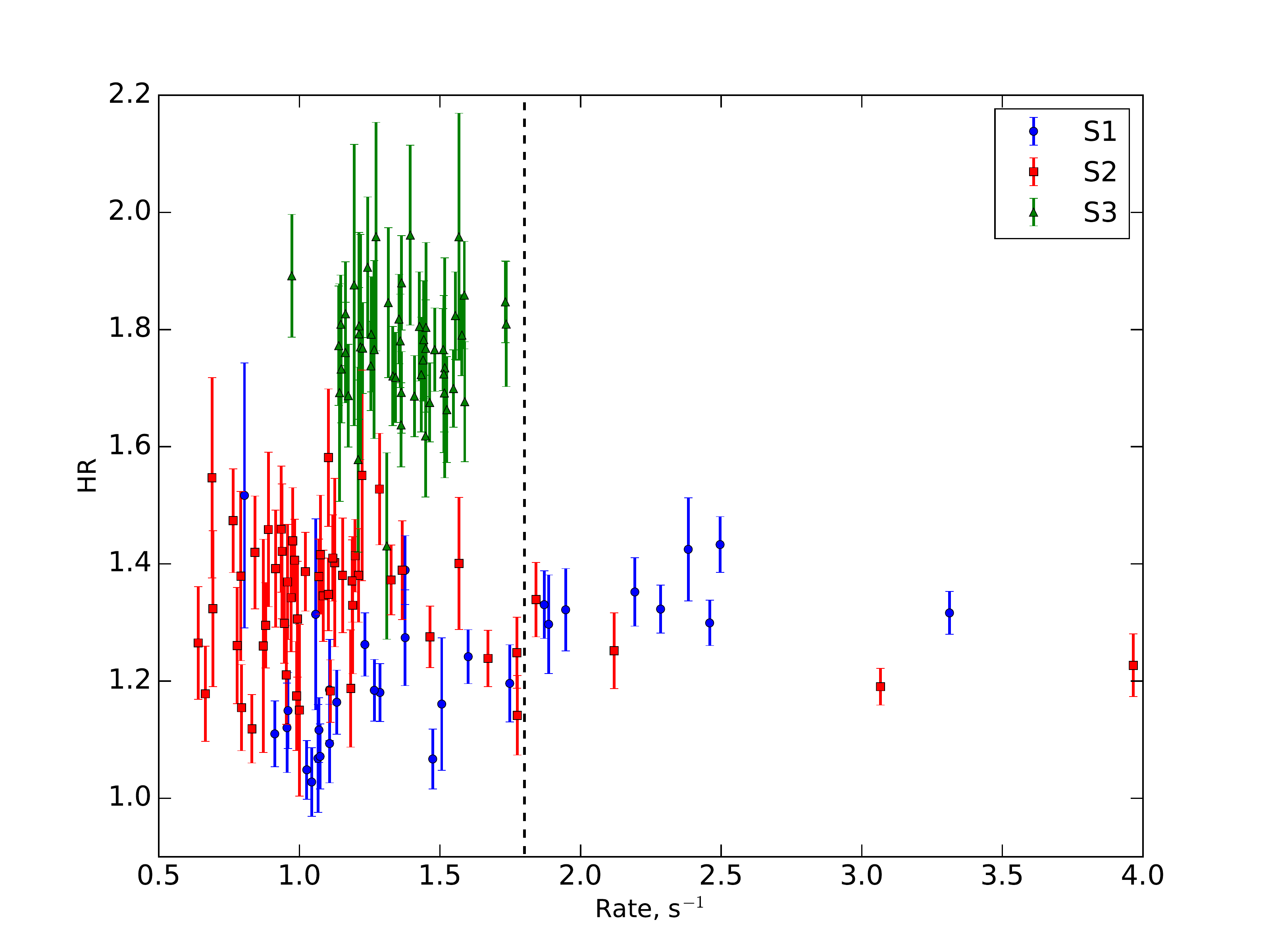}
\caption{HRs versus intensity plot for the all three \textit{Suzaku}  observations with a bin time of 2 ks. Count rates correspond to the full energy range 0.3-10.0 keV, and HR is defined as ratio between hard (2.0-10.0 keV) and soft (0.3-2.0 keV) energy bands. Dashed line corresponds to the count rate of 1.8 cts/s used to split S1 and S2 observations, see text for details.}\label{fig_hr_suz}
\end{figure}

Previous studies of the source indicate   hardening of the source spectrum with the rise of the luminosity, both within the single observation and on a longer time scale  \citep{sidoli06,smith09}. As shown in Figure \ref{fig_hr_suz} we clearly see this dependence in the case of the S1 observation,   while in the S2 observation we don't see any hint of spectral hardening during the flares. There were no flares during the S3 observation, but in general it turned out to be much harder than the S1 and S2 observations.

In spectral analysis we have used data only from the front-illuminated CCDs (XIS-0 and XIS-3), which have a larger effective area at high energies, and have almost identical properties, so that it is possible to co-add the data to improve photon statistics. For spectral fitting we used a single power law model with photoelectric absorption (\texttt{phabs*powerlaw XSPEC} model). This model provides a good fit for all three \szk\ observations, see Table~\ref{tab_suz_fit}. Absorption column densities and photon indexes for S1 and S2 are consistent with the previous results of \xmm\ and \textit{Chandra} observations \citep{sidoli06,albert08,Anderhub09}. For the S3 case (near the periastron), we got a higher value of column density and a lower value for the index. This value of the column density is in agreement with the results obtained by \textit{Chandra} in spring 2006 \citep{paredes07}. The spectral slope varies from 1.6 near periastron to  1.8 -- 1.9 near apastron, while column density changes from 0.6 to 0.5.

\begin{table*}
\centering
\caption{\textit{Suzaku} results of spectral fitting with absorbed power law model. In lines marked with (+PIN) we give the result of simultaneous fitting of XIS and PIN data. Value of  N$_{H}$ is given in 10$^{22}$ cm$^{-2}$, and values of absorbed and unabsorbed fluxes are given in 10$^{-11}$ erg cm$^{-2}$ s$^{-1}$. For the S3 observation the PIN calibration factor is found to be 1.15$\pm$0.11, as expected from the standard calibration, while for S1 and S2   observations the fitted values turned out to be too high, 1.74$\pm$0.20 and 2.63$\pm$0.23. See text for more details.  }\label{tab_suz_fit}
% \begin{tabular}{c@{ }c@{ }c@{ }c@{ }c@{ }c@{ }c@{ }c@{ }}
\begin{tabular}{cccccccc}
\hline
Ref & N$_{H}$ & $\Gamma$ & Flux$^{ abs}_{1-10 keV}$ &Flux$_{1-10 keV}$& $\chi^2_{red} (d.o.f.)$& RMS$_{1ks}$\\

\hline

S1 & 0.50$\pm$0.01 & 1.86$\pm$0.01 & 1.17$\pm$0.01 &1.38$\pm$0.01 &0.98 (375) & 0.66$\pm$0.07 \\
S1(+PIN) & 0.49$\pm$0.01 & 1.86$\pm$0.01 & 1.18$\pm$0.01 &1.38$\pm$0.01& 0.98 (381) & \\
%S1a & 0.46$\pm$0.02 & 1.90$\pm$0.03 & 0.79$\pm$0.01&0.90$\pm$0.01 & 1.03 (456) \\
%S1b & 0.52$\pm$0.02 & 1.82$\pm$0.02 & 2.13$\pm$0.02 &2.44$\pm$0.01& 0.96 (533) \\
S1a & 0.47$\pm$0.03 & 1.97$\pm$0.04 & 0.87$\pm$0.02&1.09$\pm$0.02 & 1.13 (484) \\
S1b & 0.52$\pm$0.02 & 1.88$\pm$0.03 & 2.03$\pm$0.02 &2.54$\pm$0.02& 1.06 (553) \\
S2 & 0.50$\pm$0.01 & 1.80$\pm$0.02 & 0.92$\pm$0.01&1.08$\pm$0.01 & 0.98 (437) & 0.57$\pm$0.08\\
S2(+PIN) & 0.49$\pm$0.01 & 1.80$\pm$0.01 & 0.92$\pm$0.01 &1.08$\pm$0.01& 0.98 (444)\\
S2a & 0.53$\pm$0.02 & 1.96$\pm$0.03 & 0.98$\pm$0.01 & 1.19$\pm$0.01 & 0.99 (471) \\
S2b & 0.49$\pm$0.01 & 1.96$\pm$0.02 & 1.55$\pm$0.01 & 1.88$\pm$0.01 & 0.98 (400) \\
S3 & 0.62$\pm$0.01 & 1.63$\pm$0.01 & 1.21$\pm$0.01 &1.43$\pm$0.01& 1.04 (559) & 0.17$\pm$0.02\\
S3(+PIN) & 0.62$\pm$0.01 & 1.62$\pm$0.01 & 1.21$\pm$0.01 &1.43$\pm$0.01& 1.04 (575)\\
\hline
\end{tabular}
\end{table*}

\begin{figure}
\centering
\includegraphics[width=0.9\columnwidth]{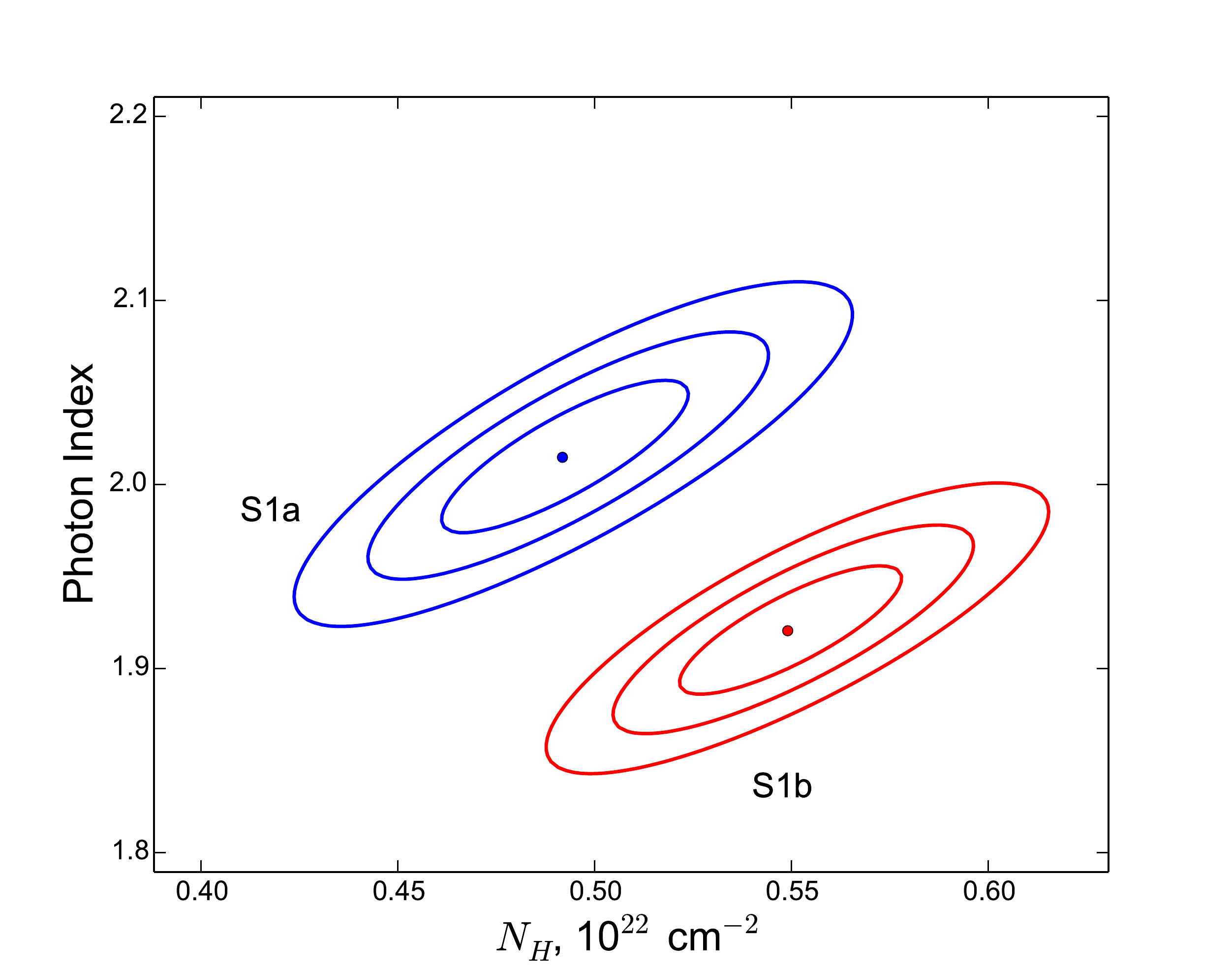}
\includegraphics[width=0.9\columnwidth]{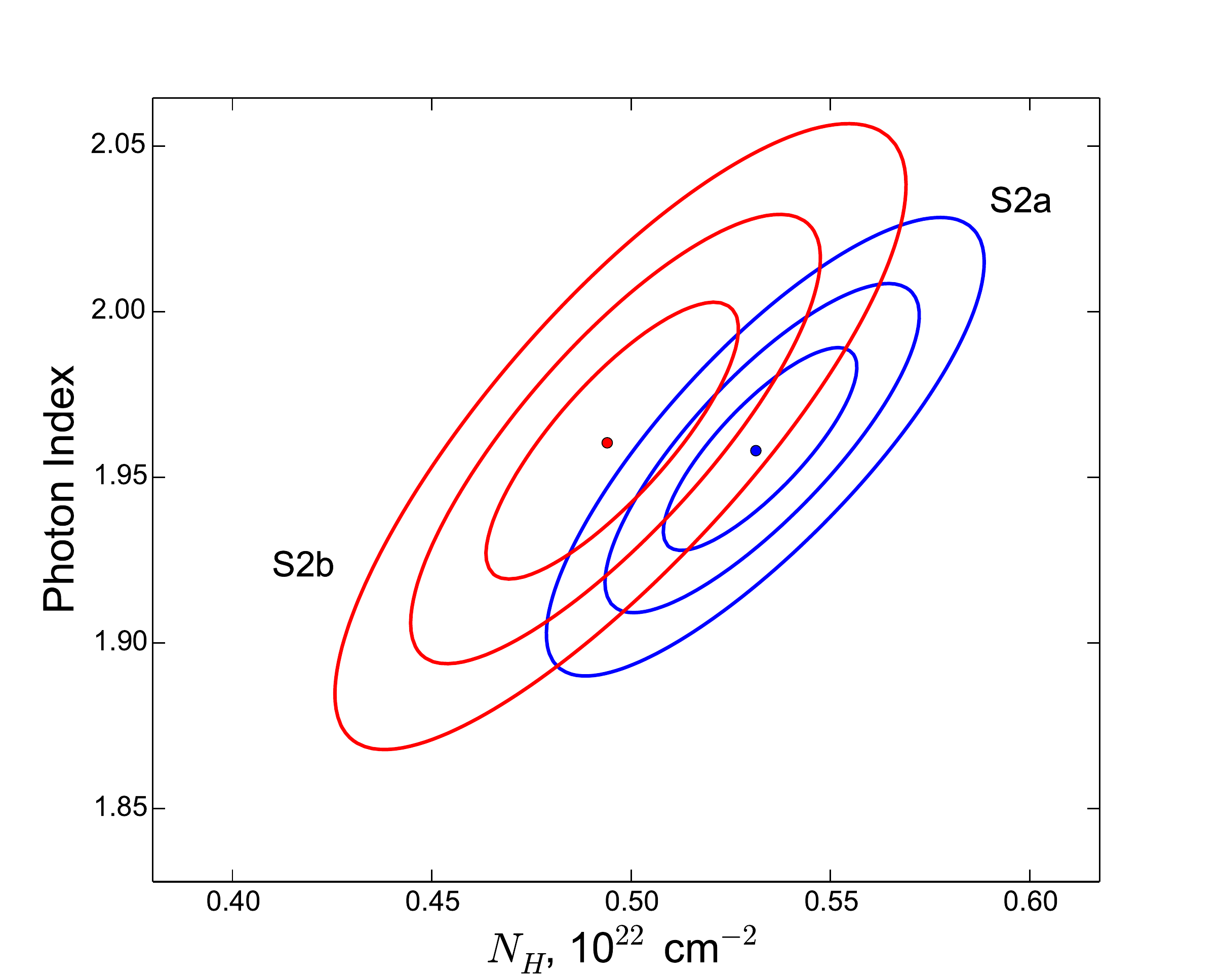}
\caption{1, 2 and 3$\sigma$ confidence contour plots of the column density $N_H$ versus photon spectral index $\Gamma$  for a power-law fit to S1a and S1b observations (top panel) and S2a and S2b (bottom panel).} 
\label{fig_con_split_suz}
\end{figure}

To compare the spectral shape of the source in and out of the flare we split S1 observation into two parts with total count rates below  (S1a) and above (S1b) 1.8 cts/s.  Spectral characteristics of S1a and S1b  turned out to be quite different indeed, see Table \ref{tab_suz_fit} and upper panel of Figure \ref{fig_con_split_suz}. A similar split for the S2 observation  (below and above 1.8 cts/s) doesn't show notable change of the spectral index, but instead shows a variation of the column density at a 1 $\sigma$ level, see Table \ref{tab_suz_fit} and bottom panel of Figure \ref{fig_con_split_suz}. 

\begin{figure}
\centering
\includegraphics[width=0.9\columnwidth]{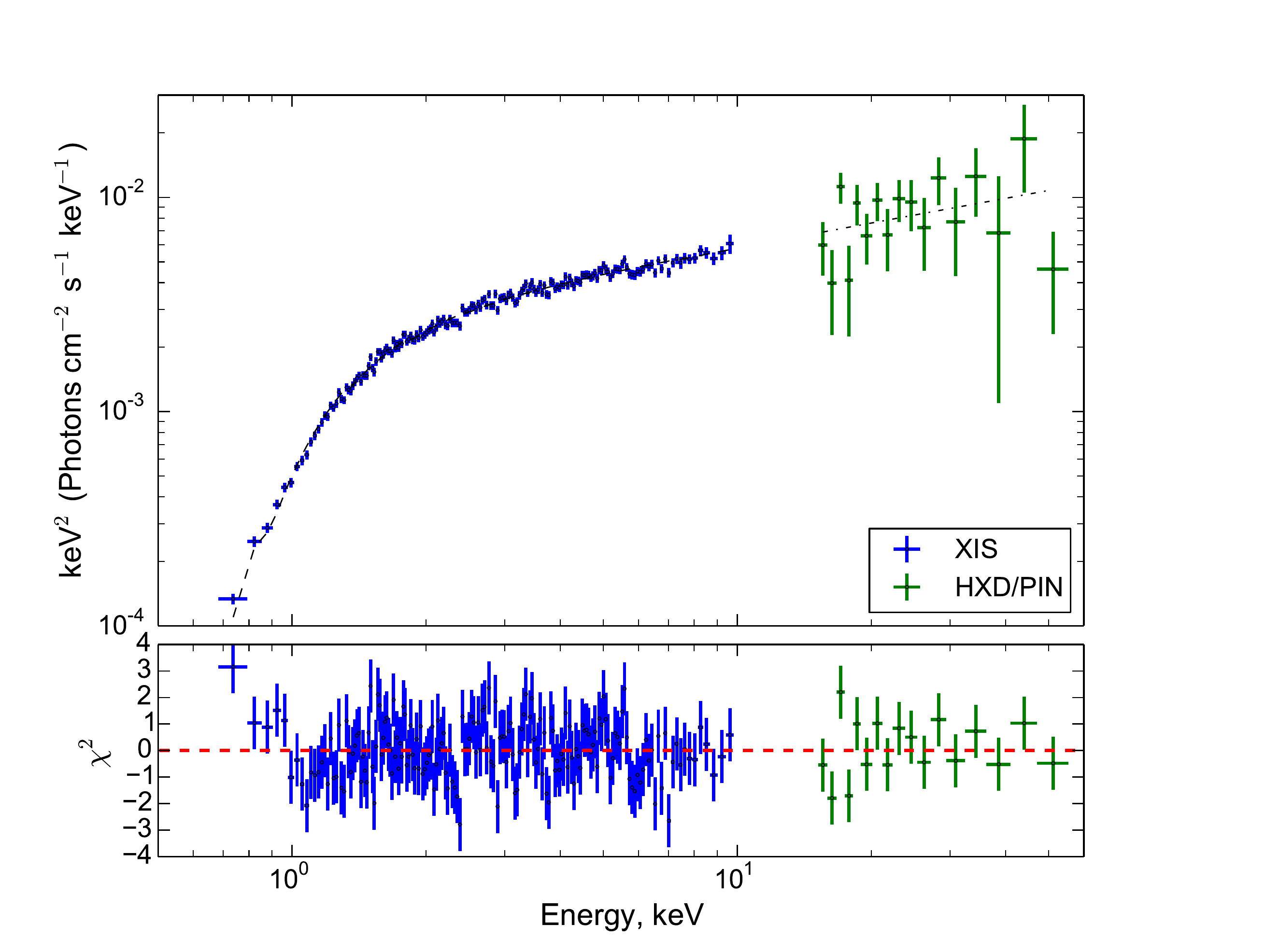}
\caption{\textit{Suzaku} spectrum of  S3 observation fitted with absorbed power law model.  }\label{fig_spec_suz}
\end{figure}

For the HXD/PIN data,  we have used standard event reprocessing and screening criteria. The tuned background files were chosen to determine the HXD/PIN non-X-ray background. To extract the source spectrum of \lsi\ we used the \texttt{hxdpinxbpi} tool. For the S3 observation HXD/PIN, data (corrected using the standard cross calibration factor of 1.15 ) lies on the continuation of the power law model   used to describe S3 XIS data, see Figure \ref{fig_spec_suz}.  For S1 and S2 observations, however,  HXD/PIN points lie well above the model derived from XIS data (cross calibration factor 1.74 for S1 and 2.63 for S2 are needed for a plausible fit). It turned out that during the S1 and S2 observations the temperature of HXD was too high, which resulted in a very high instrumental background leading to telemetry saturation and dead-time of about 50\% (typically the dead-time is $<$ 10\%). In addition to that during that week there was almost no earth occultation data. Such a situation  is not well treated in the current background generation,  so that the  background was underestimated, leading to extremely high values of the calibration factor.

\subsection{\xmm\ observations}

\begin{table*}
\caption{The log of XMM-Newton observations}\label{tab_xmm_log}
\begin{tabular}{cccccccc}
\hline
Obs. date & ObsID & Ref & MJD &$\phi$& $\Phi$ & Exp.time & RMS$_{1ks}$ \\
& & & & & & ks& \\
\hline
05-02-2002 & 0112430101& X1 & 52310.072 &  0.552 & 0.3652& 6.40   & 0.13$\pm$0.02 \\
10-02-2002 & 0112430102& X2 & 52315.461 & 0.756 & 0.3684 & 6.40   & 0.65$\pm$0.01 \\
17-02-2002 & 0112430103& X3 & 52322.179 & 0.009 & 0.3724 & 6.40   & 0.39$\pm$0.01 \\
21-02-2002 & 0112430201& X4 & 52326.639 & 0.178 & 0.3751 & 7.49   & 0.22$\pm$0.01 \\
16-09-2002 & 0112430401& X5 & 52533.099 & 0.970 & 0.4990 & 6.46   & 0.10$\pm$0.02 \\
27-01-2005 & 0207260101& X6 & 53397.736 & 0.602 & 0.0176 & 50.41  & 0.60$\pm$0.01 \\
04-09-2007 & 0505980801& X7 & 54347.075 & 0.432 & 0.5871 & 17.41  & 0.28$\pm$0.01 \\
06-09-2007 & 0505980901& X8 & 54349.063 & 0.507 & 0.5883 & 13.41  & 0.18$\pm$0.01\\
07-09-2007 & 0505981001& X9 & 54350.092 & 0.546 & 0.5890 & 18.55  & 0.25$\pm$0.01\\
08-09-2007 & 0505981101& X10 & 54351.060 & 0.582 & 0.5895 & 17.41 & 0.33$\pm$0.01 \\
09-09-2007 & 0505981201& X11 & 54352.060 & 0.620 & 0.5901 & 15.01 & 0.14$\pm$0.01 \\
10-09-2007 & 0505981301& X12 & 54353.058 & 0.658 & 0.5907 & 16.87 & 0.09$\pm$0.01 \\
11-09-2007 & 0505981401& X13 & 54354.060 & 0.703 & 0.5913 & 14.51 & 0.27$\pm$0.01\\
\hline
\end{tabular}
\end{table*}

\begin{figure}
\includegraphics[width=\columnwidth]{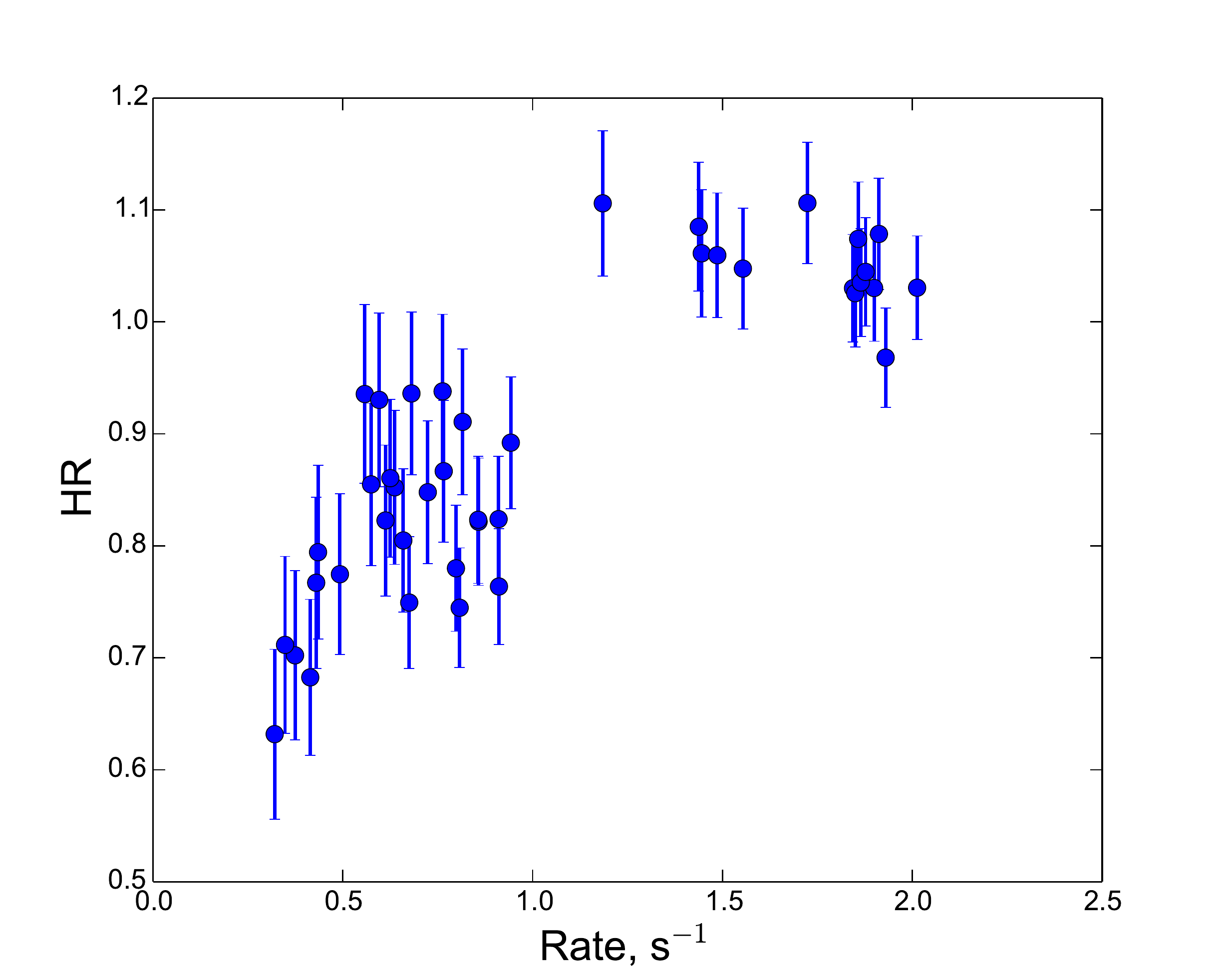}
\caption{The dependence of HR (ratio of count rate in the 2 -- 12 keV energy range to count rate in the 0.3 -- 2 keV energy range) on the count rate in the full energy band (0.3-12.0 keV) as observed by \xmm\ during  X6 PN observation. Time binning is 1 ks.}\label{fig_lc_xmm}
\end{figure}

\begin{figure*}
\includegraphics[width=0.47\linewidth]{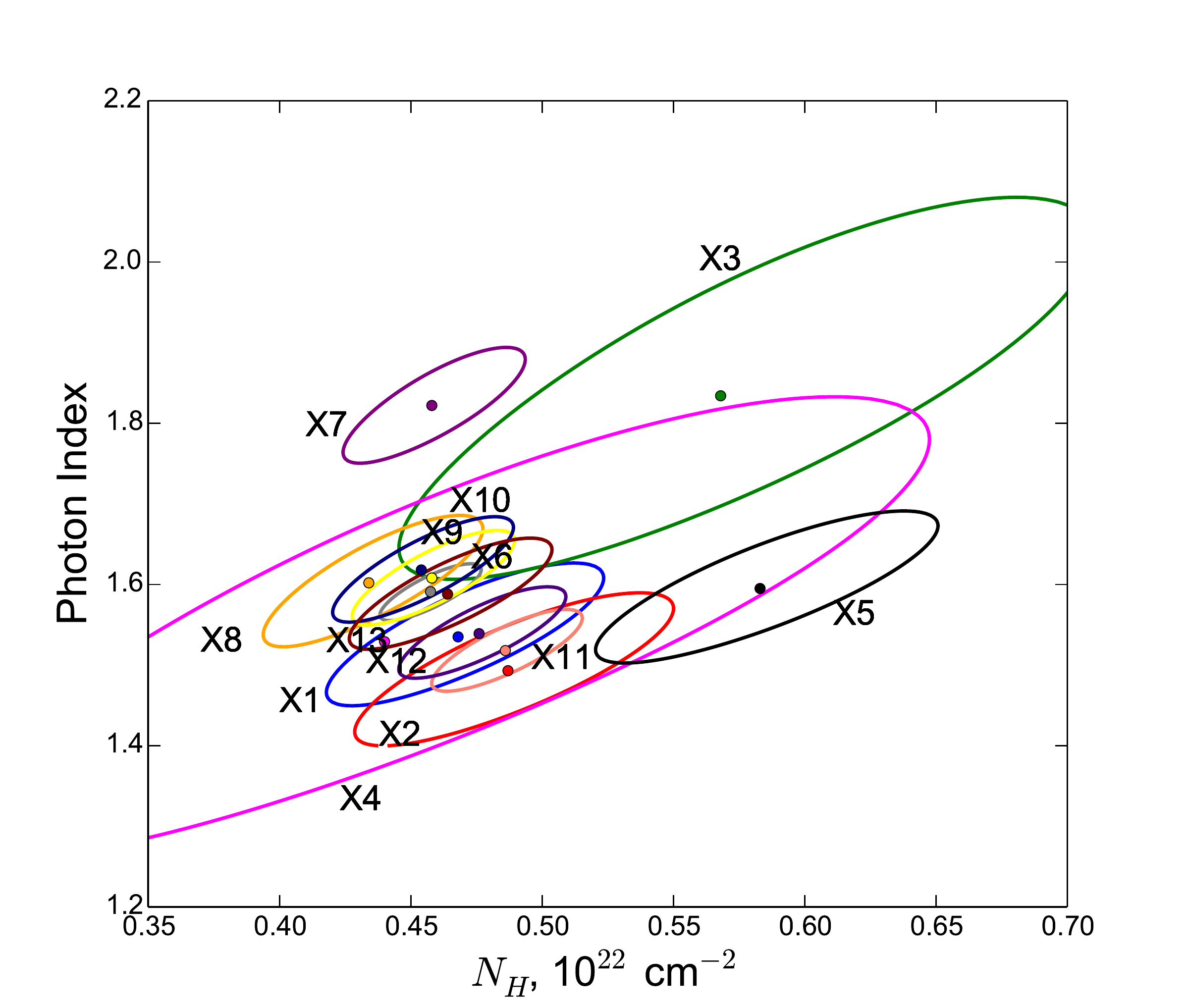}
\includegraphics[width=0.47\linewidth]{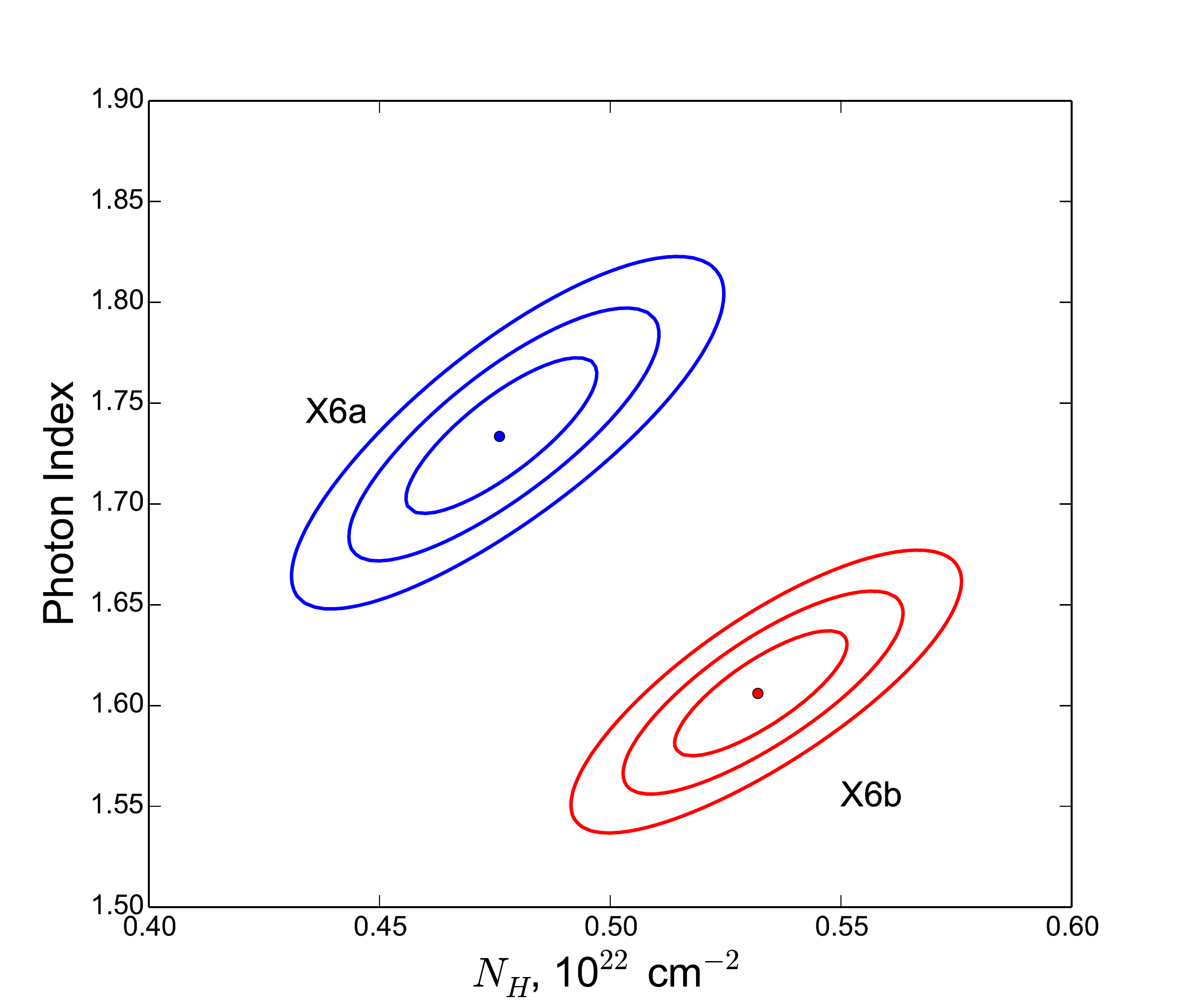}
\caption{ Contour plots of the column density $N_H$ versus photon spectral index $\Gamma$  for a power-law fit to all \xmm\ observations (left panel, only $3\sigma$ contours are shown) and X6a and X6b (right panel, 1, 2, and 3 $\sigma$ contours are shown). }\label{fig_cont_xmm}
\end{figure*}

The \xmm\ Observatory consists of three  X-ray telescopes. Two of these telescopes use MOS CCDs \citep{herder01} and one uses a PN CCD \citep{struder01}. \xmm\ observed \lsi\ 13 times during the last decade (see Tab.~\ref{tab_xmm_log} for details). These data were already presented in \cite{chernyakova06,sidoli06,Anderhub09}. Here we reanalyzed them using the latest \xmm\ Science Analysis Software v.14.0.0 (SAS). Known hot pixels and electronic noise were removed, and data were cleaned from the influence of the soft proton flares. In our analysis, we have used only PN data to extract all the products.   The source spectra were extracted from a circular area of 36 arc seconds radius, and the background spectra were obtained from a nearby, source free region of the same radius. Spectra were rebinned to have at least 30 counts per bin. The response matrix files (RMFs) and auxiliary response files (ARFs) were extracted using RMFGEN and ARFGEN tools, respectively. A single power law model with photoelectric absorption (\texttt{phabs*powerlaw XSPEC} model) provides a good fit for all \xmm\ observations, see Table~\ref{tab_xmm_res}. It turned out that both spectral index and hydrogen column density varies along the orbit, as is illustrated on the left panel of Figure~\ref{fig_cont_xmm}, where we show  $3\sigma$ contour plots. It is clearly seen that for e.g.  X5 and X7 observations both slope and  column density are different at more than  $3\sigma$  confidence level.
During the longest \xmm\ observation, X6, \cite{sidoli06} has observed a sharp softening of the source spectrum in the middle of the observation. The dependence of the hardness ratio on the source count rate is shown in Figure \ref{fig_lc_xmm}. In our spectral analysis, similar to S1 and S2 \szk\ observations, we split X6  into X6a, with count rate in 0.3 -- 12 keV below 1.1 cts/s and X6b with the count rate above 1.1 cts/s. Hardening of the  source spectrum with the rise of the source flux is clearly seen (see Table \ref{tab_xmm_res} and right panel of Figure \ref{fig_cont_xmm}).

\begin{table}
\centering
\caption{Best fit results of the spectral analysis for \xmm\ observations of \lsi. The flux values are in units of 10$^{-11}$ erg/cm$^{2}$/s, while absorption is in units of $10^{22}$ cm$^{-2}$.}\label{tab_xmm_res}
\begin{tabular}{cccccl}
\hline
Ref & N$_{H}$ & $\Gamma$ & Flux$^{abs}_{1-10 keV}$ & Flux$_{1-10 keV}$ &$\chi^{2}_{red}$ (d.o.f.) \\
\hline

X1 & 0.47$\pm$0.01 & 1.53$\pm$0.02 & 1.59$\pm$0.02 & 1.84$\pm$0.01 &1.07 (253) \\
X2 & 0.49$\pm$0.02 & 1.49$\pm$0.03 & 1.47$\pm$0.02 & 1.70$\pm$0.02 &1.07 (223) \\
X3 & 0.57$\pm$0.04 & 1.83$\pm$0.07 & 0.76$\pm$0.03 & 0.93$\pm$0.02 &0.98 (95) \\
X4 & 0.43$\pm$0.05 & 1.52$\pm$0.08 & 0.52$\pm$0.02 & 0.62$\pm$0.03 &1.01 (87) \\
X5 & 0.57$\pm$0.02 & 1.57$\pm$0.03 & 1.48$\pm$0.04 & 1.72$\pm$0.02 &0.93 (251) \\
X6 & 0.45$\pm$0.01 & 1.58$\pm$0.01 & 0.98$\pm$0.01 &1.11$\pm$0.01 & 1.14 (1074) \\
X6a & 0.44$\pm$ 0.01 & 1.73$\pm$0.02 & 0.50$\pm$0.01&0.57$\pm$0.01 & 1.05 (223) \\
X6b & 0.50$\pm$0.01 & 1.61$\pm$0.02 & 1.44$\pm$0.02 &1.67$\pm$0.01 & 1.12 (289) \\
X7 & 0.45$\pm$0.01 & 1.82$\pm$0.02 & 0.83$\pm$0.09 &0.95$\pm$0.01& 1.07 (357)\\
X8 & 0.44$\pm$0.01 & 1.61$\pm$0.02 & 0.86$\pm$0.01 &0.96$\pm$0.01& 1.09 (348) \\
X9 & 0.46$\pm$0.01 & 1.60$\pm$0.02 & 0.91$\pm$0.01 &1.03$\pm$0.01& 1.29 (505) \\
X10 & 0.45$\pm$0.01 & 1.62$\pm$0.02 & 0.92$\pm$0.01 &1.02$\pm$0.01& 0.97 (500) \\
X11 & 0.48$\pm$0.01 & 1.51$\pm$0.01 & 1.62$\pm$0.02 &1.80$\pm$0.01& 1.15 (634) \\
X12 & 0.47$\pm$0.01 & 1.53$\pm$0.02 & 1.31$\pm$0.01 &1.44$\pm$0.01& 1.08 (513) \\
X13 & 0.46$\pm$0.01 & 1.58$\pm$0.02 & 0.88$\pm$0.01 &0.98$\pm$0.01& 1.10 (376) \\

\hline
\end{tabular}
\end{table}

\subsection{\textit{Chandra} observations}
\lsi\ has been observed by the \textit{Chandra} X-ray observatory 3 times to date (see Table~\ref{tab_ch_log}). These data were already presented by \cite{paredes07, albert08, Rea10}. For consistency we reanalyze these data with the latest version of software available, CIAO v.4.3 software package and  CalDB v.4.3.0. To extract source and background spectra with RMF and ARF we used the {\tt specextract} tool. The source spectra were extracted from a circular area around the source with a 3 arc seconds radius, while the background spectra were made from a nearby region with no source and with the same 3 arc second radius. The source spectra were grouped to have at least 30 counts per bin. The fitting of spectra was performed in \texttt{XSPEC} environment with \texttt{phabs*powerlaw} model. Following the work of \cite{paredes07} we have also added the pileup model to the absorbed power law for the first \textit{Chandra} observation C1. For this observation we 
fixed the event pileup fraction parameter $f$ to 0.95, and the  grade migration parameter was found to be $\alpha=0.33\pm 0.09$.
The results of the fitting are shown in Table~\ref{tab_ch_res}. 

\begin{table}
\caption{The log of \textit{Chandra} data of \lsi}\label{tab_ch_log}
\begin{tabular}{c@{ }c@{ }c@{ }c@{ }c@{ }c@{ }cc@{ }}
\hline
Obs. date & ObsID & Ref & MJD & Instr. & Exp.time & $\phi$ & $\Phi$ \\
 & & & d & & ks & & \\
\hline
07-04-2006 & 6585 & C1 & 53832.922 & ACIS-I & 49.73 & 0.027 & 0.2787 \\
25-10-2006 & 8273 & C2 & 54033.925 & ACIS-I & 20.03 & 0.613 & 0.3993 \\
14-11-2008 & 10052 & C3 & 54784.460 & ACIS-S & 95.67 & 0.940 & 0.8495 \\
\hline
\end{tabular}
\end{table}

\begin{table}
\caption{The best fit parameters of \textit{Chandra} data of \lsi. Value of  N$_{H}$ is given in 10$^{22}$ cm$^{-2}$, and values of absorbed and unabsorbed fluxes are given in 10$^{-11}$ erg/cm$^{2}$/s} \label{tab_ch_res}
\begin{tabular}{c@{ }c@{ }c@{ }c@{ }c@{ }c@{ }c@{ }c@{ }c@{ }}
\hline
Ref & N$_{H}$ & $\Gamma$ & F$^{ abs}_{1-10 keV}$ & F$_{1-10 keV}$&$\chi^{2}$/dof & RMS$_{1ks}$\\

\hline
C1 & 0.65$\pm$0.045 & 1.30$\pm$0.08 & 0.50$\pm$0.01&0.60$\pm$ 0.01 & 1.19/293 & 0.36$\pm$0.01\\
C2 & 0.56$\pm$0.02 & 1.55$\pm$0.02 & 2.32$\pm$0.04 &2.66$\pm$0.02 & 1.03/329 & 0.52$\pm$0.01\\
C3 & 0.59$\pm$0.01 & 1.78$\pm$0.01 & 0.82$\pm$0.01 &0.98$\pm$ 0.01& 1.09/460 & 0.19$\pm$0.01\\
\hline
\end{tabular}

\end{table}

\subsection{\sw observations}
Observations performed by the \sw observatory during the last few years give a unique chance to monitor the behaviour of \lsi on a range of time scales. In this work we used data spanning over more than 4 years between 2010 and 2014.  The data reduction was done using tools and packages available in {\tt FTOOLS/HEASOFT 6.16}. \lsi was observed both in Photon Counting (PC) and Windowed Timing (WT) modes. After initial cleaning of events using {\tt  xrtpipeline} with standard parameters we selected 164 observations. 
 Further analysis was
performed following \cite{evans2009}. In particular, 
in the PC mode
the source extraction region was a circle with radius from 5 to 30 pixels
depending on the count rate 
\citep{evans2009}; in the WT mode the radius of the source
extraction region was 25 pixels. The background was collected over the
annular region with an inner (outer) radius of 60 (110) pixels in both
observational modes. The count rate from the source was too low to pile up
the detector in all observations.

\begin{figure}
\centering
\includegraphics[width=\linewidth]{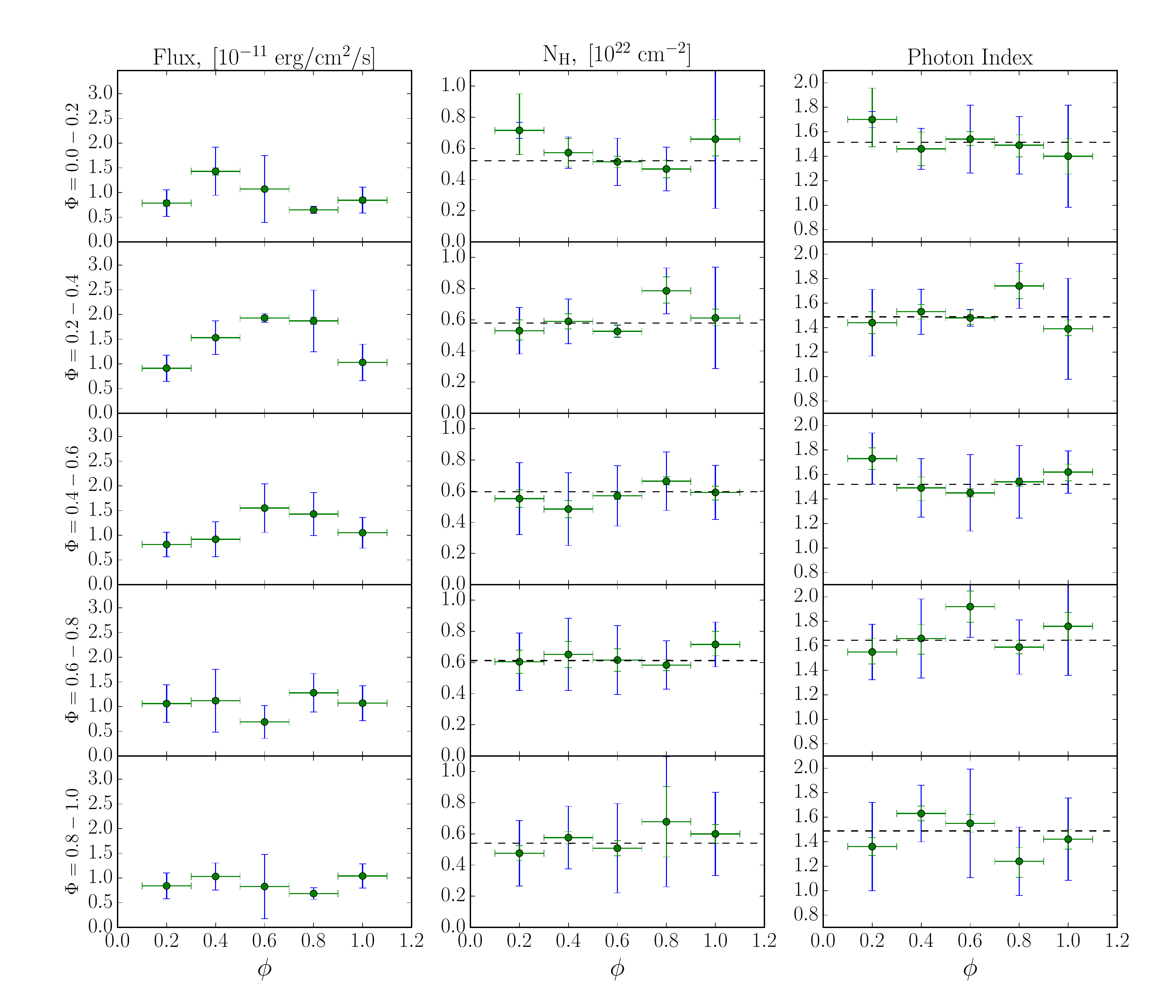}
\caption{Dependence of flux, column density and spectral index on the orbital phase at different superorbital phases as measured by \sw. }
\label{swift_op_sop}
\end{figure}

Since the statistics of the data is poor, the spectra were grouped to have at least 1 count bin$^{-1}$ using the FTOOLS {\tt grppha} tool. The spectral analysis was performed in \texttt{XSPEC} environment and the spectra were fitted at 0.5 -- 10.0 keV energy band. The errors reported in this work are purely statistical and correspond to a 1$\sigma$ confidence level. For most of the observations the statistic was too poor to draw any conclusions from a single observation.  Thus we selected observations from similar orbital and superorbital phases and fitted them together. The results are given in Table \ref{tab_swift_res_group} and Figure \ref{swift_op_sop}, which show the 1.0 - 10.0 keV X-ray flux, column density and photon index from the source as a function of the orbital and superorbital phases. Note that each point in this figure is a combination of several closely spaced observations. Green error bars correspond to the 1$\sigma$ errors of the spectral fit, while blue error bars were calculated  taking into account the spread of spectral parameters of the individual observations (following the equation for RMS given in section 2.1). These latter error bars represent the potential variability of the source in each of the orbital phase bins; however, they should be treated with caution given the poor statistics of the single \sw measurements.
One can see that in agreement with previous findings \citep{chernyakova12,li12}  the flux is maximal during the 
 0.2-0.4 superorbital phase, with a peak becoming wider and moving from orbital phase $\phi \sim 0.4$ at superorbital phase $\Phi =$ 0 - 0.2 to orbital phase $\phi \sim 0.8$ at  superorbital phase $\Phi = $ 0.4 - 0.6. During the superorbital phases $\Phi = $ 0.6 - 0 the peak disappears, or become too narrow to be detected with such wide bins.  On the contrary, the values of  column density and photon index are consistent with a constant value, the same for  all  orbital and superorbital phases (the averaged value of column density is equal to  $<N_H>=0.55 \pm 0.01$, RMS=0.24; the averaged value of spectral index is equal to $<\Gamma >=  1.54  \pm 0.02$, RMS=0.32). The evolution of the source X-ray flux with the orbital phase shows X-ray emission along all the phases, although with a difference of a factor 3 in the flux level between two states, with a ``high'' state preferentially found in the phase range 0.4-1.0.

\begin{table}
\caption{Best fit results of the spectral analysis for \sw observations of \lsi sampled along different orbital and superorbital phases. The flux values are in units of 10$^{-11}$ erg/cm$^{2}$/s, while absorption in units of $10^{22}$ cm$^{-2}$. ``C-stat'' represents the value of the Cash statistic${{}^1}$, estimated by XSpec.}
\label{tab_swift_res_group}
\begin{tabular}{ccccccc}
\hline
$\phi$ &  $\Phi$ &  F$_{1-10 keV}$ &  $N_H$   &  $\Gamma$  & $C-stat$ & $\chi^2$/(d.o.f)\\
\hline
0.9-1.1 & 0.8-1.0 & 1.04$\pm$0.03&  0.60$\pm$0.06 & 1.42$\pm$0.08 &     430.71 & 1.22/484\\
0.7-0.9 & 0.8-1.0 & 0.68$\pm$0.01&  0.68$\pm$0.22 &   1.24$\pm$0.13 &   325.39 & 1.00/408\\
0.5-0.7 & 0.8-1.0 & 0.83$\pm$0.02&  0.51$\pm$0.05 &   1.55$\pm$0.07 &   442.72 & 1.25/508\\
0.3-0.5 & 0.8-1.0 & 1.03$\pm$0.02&  0.58$\pm$0.04 &   1.63$\pm$0.06 &   447.92 & 1.04/570\\
0.1-0.3 & 0.8-1.0 & 0.84$\pm$0.03&  0.48$\pm$0.05 &   1.36$\pm$0.07 &   456.42 & 1.28/493\\
0.9-1.1 & 0.6-0.8 & 1.07$\pm$0.04&  0.72$\pm$0.08 &   1.76$\pm$0.11 &   278.34 & 0.92/379\\
0.7-0.9 & 0.6-0.8 & 1.28$\pm$0.03&  0.58$\pm$0.04 &   1.59$\pm$0.06 &   568.44 & 1.35/598\\
0.5-0.7 & 0.6-0.8 & 0.69$\pm$0.03&  0.62$\pm$0.07 &   1.92$\pm$0.13 &   259.08 & 1.06/324\\
0.3-0.5 & 0.6-0.8 & 1.12$\pm$0.05&  0.65$\pm$0.08 &   1.66$\pm$0.13 &   303.18 & 1.14/356\\
0.1-0.3 & 0.6-0.8 & 1.06$\pm$0.04&  0.60$\pm$0.07 &   1.55$\pm$0.11 &   307.78 & 0.96/411\\
0.9-1.1 & 0.4-0.6 & 1.05$\pm$0.02&  0.59$\pm$0.05 &   1.62$\pm$0.07 &   495.65 & 1.18/559\\
0.7-0.9 & 0.4-0.6 & 1.43$\pm$0.02&  0.66$\pm$0.03 &   1.54$\pm$0.04 &   779.45 & 1.40/723\\
0.5-0.7 & 0.4-0.6 & 1.55$\pm$0.02&  0.57$\pm$0.03 &   1.45$\pm$0.03 &   750.02 & 1.32/747\\
0.3-0.5 & 0.4-0.6 & 0.92$\pm$0.04&  0.48$\pm$0.05 &   1.49$\pm$0.10 &   370.82 & 1.38/408\\
0.1-0.3 & 0.4-0.6 & 0.81$\pm$0.03&  0.55$\pm$0.06 &   1.73$\pm$0.09 &   363.66 & 1.28/434\\
0.9-1.1 & 0.2-0.4 & 1.03$\pm$0.02&  0.61$\pm$0.06 &   1.39$\pm$0.07 &   585.15 & 1.36/591\\
0.7-0.9 & 0.2-0.4 & 1.87$\pm$0.08&  0.79$\pm$0.08 &   1.78$\pm$0.12 &   301.17 & 1.02/378\\
0.5-0.7 & 0.2-0.4 & 1.93$\pm$0.05&  0.53$\pm$0.04 &   1.48$\pm$0.06 &   497.10 & 1.12/577\\
0.3-0.5 & 0.2-0.4 & 1.53$\pm$0.04&  0.59$\pm$0.05 &   1.53$\pm$0.06 &   524.21 & 1.33/562\\
0.1-0.3 & 0.2-0.4 & 0.91$\pm$0.04&  0.53$\pm$0.07 &   1.44$\pm$0.09 &   361.12 & 1.19/434\\
0.9-1.1 & 0.0-0.2 & 0.84$\pm$0.05&  0.66$\pm$0.11 &   1.40$\pm$0.14 &   203.29 & 0.97/250\\
0.7-0.9 & 0.0-0.2 & 0.65$\pm$0.02&  0.47$\pm$0.06 &   1.49$\pm$0.10 &   350.87 & 1.13/413\\
0.5-0.7 & 0.0-0.2 & 1.07$\pm$0.02&  0.51$\pm$0.04 &   1.54$\pm$0.06 &   584.30 & 1.32/599\\
0.3-0.5 & 0.0-0.2 & 1.43$\pm$0.07&  0.57$\pm$0.09 &   1.36$\pm$0.14 &   244.08 & 1.05/317\\
0.1-0.3 & 0.0-0.2 & 0.79$\pm$0.06&  0.72$\pm$0.23 &   1.70$\pm$0.26 &   135.98 & 0.93/180\\
\hline 
\end{tabular}
\end{table}

\footnotetext[1]{{https://heasarc.gsfc.nasa.gov/xanadu/xspec/manual/XSappendixStatistics.html, \newline http://cxc.harvard.edu/sherpa/ahelp/cstat.html}}
\setcounter{footnote}{1}

%%%%%%%%%%%%%%%%%%%%%%%%%%%%%%%%%%%%%%
\section{Variability on different time scales.}
%%%%%%%%%%%%%%%%%%%%%%%%%%%%%%%%%%%%%%

\subsection{Minimal variability time scale}
\label{sect::min_var}

% -----------------------------
% *** Suzaku power spectrum ***
\begin{figure}
  \centering
  \includegraphics[width=\columnwidth]{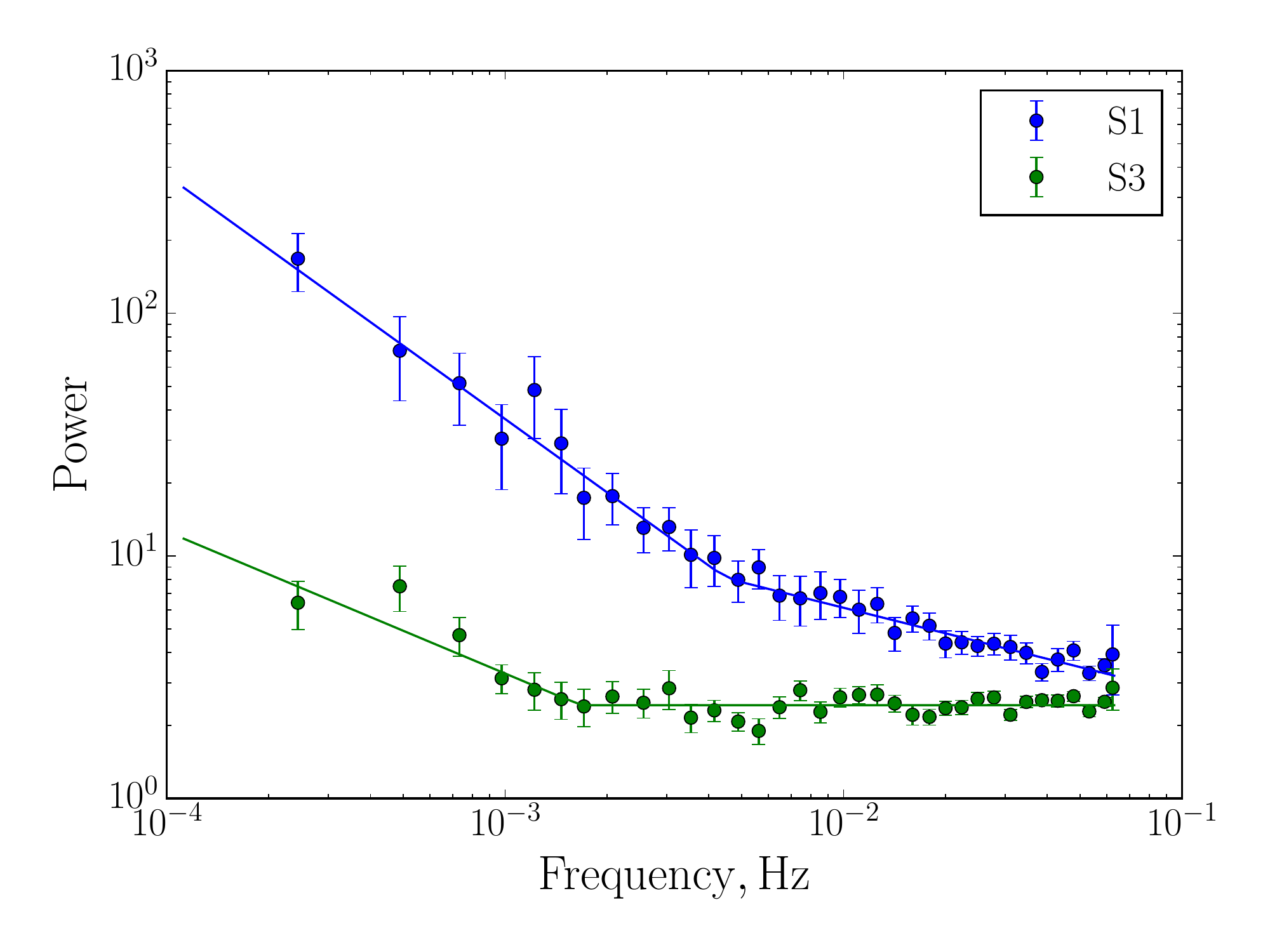}
  \caption{The FFT power spectra of the S1 and S3 observations of \lsi. The solid lines represent the corresponding fits with the broken power law. Time bin of the light curves used is 8~s.}\label{fig_ps_suz}
\end{figure}
% -----------------------------

In addition to well-established orbital and superorbital modulations, the X-ray flux of the \lsi binary system experiences much faster variability, signatures of which were found in almost all observational periods, described above. The detection of this variability allows us to address the physics of the X-ray emission region in the system through the measurement of the characteristic time scales, associated with it, as well as the overall shape of its power spectrum density  (PSD).

In order to assess the character and time scale of this variability, we analysed all available \textit{Suzaku}, \textit{Chandra} and \textit{XMM-Newton} observations; the \textit{Swift} observations turned out to be short to investigate $\sim$ks variability, detected here with other instruments, and thus were not included to the analysis. We also additionally checked, that none of the single \textit{Swift} observations shows evidences for the flux variability, being inconsistent with a constant at a more than $3\sigma$ significance level.

For each of the analysed observations we computed the power spectrum with the help of \textit{powspec} tool from the \textit{FTOOLS} package\footnote{https://heasarc.gsfc.nasa.gov/ftools/ftools\_menu.html}. The examples of the derived power spectra -- for S1 and S3 observations -- are shown in Figure~\ref{fig_ps_suz}. {For these two data sets the derived spectra exhibit clear breaks in the frequency range $10^{-3}-10^{-2}$~Hz. Still, the slope of the S1 PSD (which never becomes flat) and small uncertainties in the S3 PSD suggest that the minimal variability time scale in these data sets can be below $\sim$ 100~s.}

\begin{figure}
  \centering
  \includegraphics[width=\columnwidth]{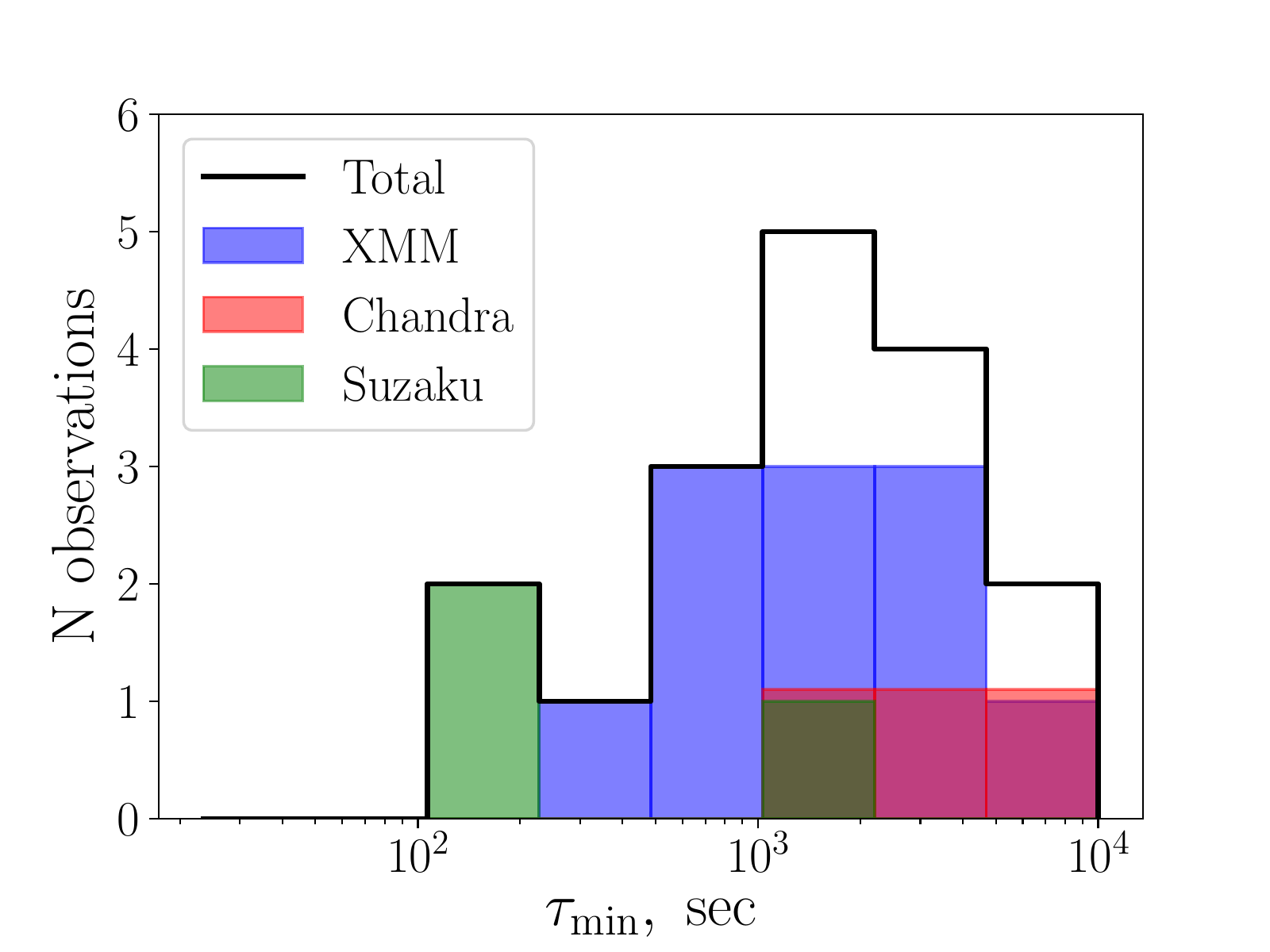}
  \caption{Distribution of the minimal variability time scales, derived with the Structure Function analysis (see Sect.~\ref{sect::min_var} for details). {Chandra points were artificially raised by 10\% to improve the visual perception.}}
  \label{fig::min_var_timescales}
\end{figure}

{However for most of the observations} large {statistical} errors prevent us from determining the break frequency at which the PSD changes it shape. Thus we have to use other means to look for both the minimal and typical variability time scales in the light curves. The minimal variability time scale was searched for by means of the Structure Function (SF) analysis as well as the scanning procedure, described in~\citet{chernyakova15}. The minimal variability time scale $\tau_{min}$ we defined as the point in SF, where it's local power law index does not equal 0 or 2 \citep[SF$\sim \tau^\alpha$ with $\alpha=0$ for white noise and $\alpha=2$ for linearly increasing flux, see][]{Me_fast_var}. The histogram of the derived values of $\tau_{min}$ is shown in Fig.~\ref{fig::min_var_timescales}.
 
% -----------------------------
% *** Suzaku power spectrum ***
% -----------------------------

The typical value of $\tau_{min}$, revealed by the SF analysis, lies in the range $\sim 200 - 1000$~s {(larger values mostly correspond to the observations where no significant variability was detected; for such data sets $\tau_{min}$ is comparable to the observation duration)}. {Still, a visual inspection of the SF shape suggests that for the C1, X4, X6 and X8 observations the SF does not reach an $\alpha=0$ plateau at the lowest time scales and rather continues a power law decline. Though this effect is not significant (i.e. does not reach the $3\sigma$ threshold), this suggests that for these observations the minimal variability time scale may be shorter that what can be resolved by our analysis - i.e. $\tau_{min} \lesssim 200$~s. For these observations, we can put an upper limit on the variability amplitude below 200~s at the level of $\Delta F/F\ \sim 15-30\%$.}

\begin{figure}
\centering
\includegraphics[width=1.1\columnwidth]{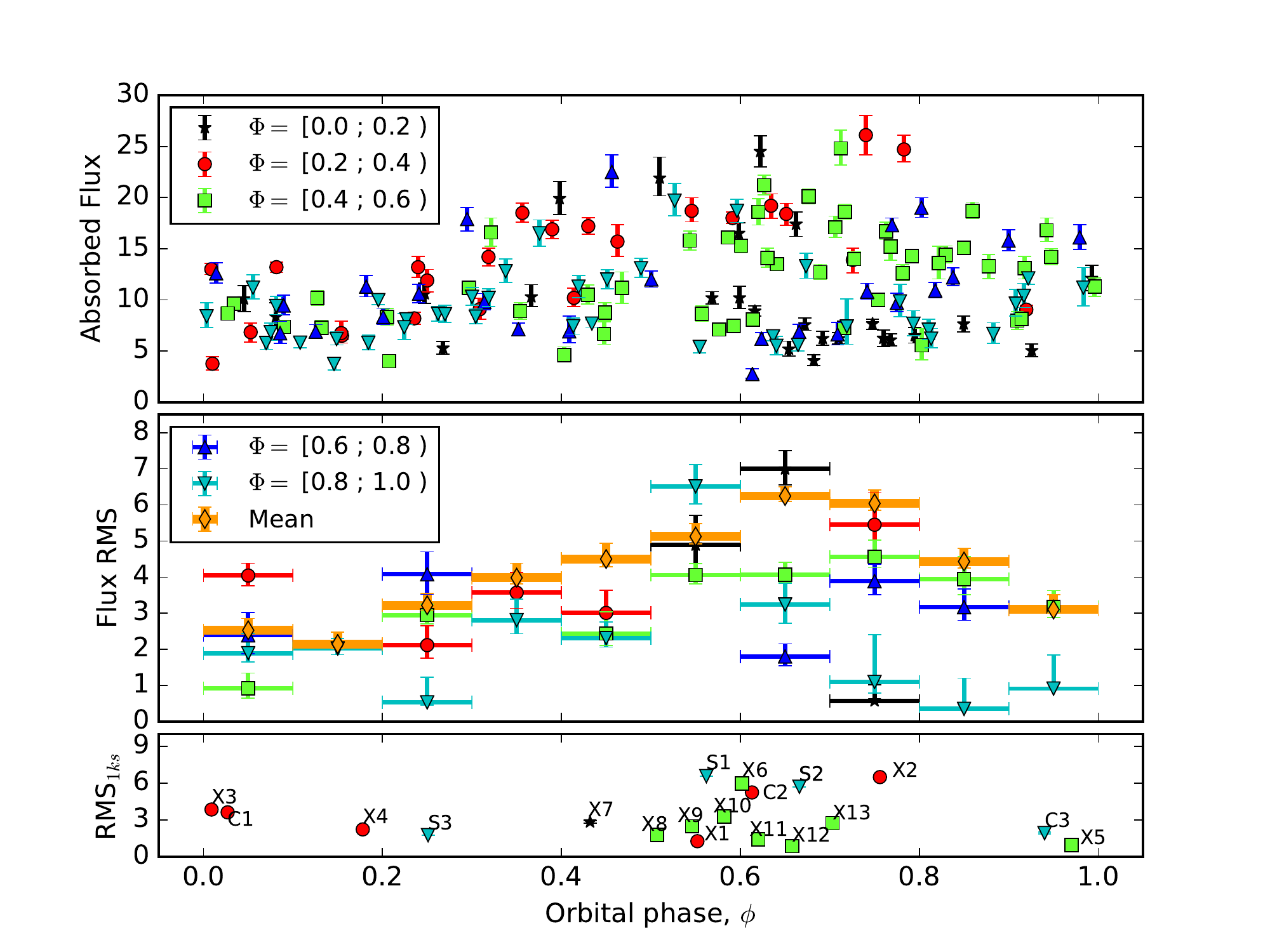}
\caption{ \textit{Top panel:} Absorbed flux (1-10~keV, in units $10^{-12}$~erg/cm$^2$/s), measured by \sw as a function of orbital phase for the specified superorbital phase ranges. \textit{Middle panel:} RMS of the shown on the top panel fluxes for the specified superorbital phase ranges.  Orange diamond points stand for the RMS at given orbital phase averaged over all superobital phases. \textit{Bottom panel:} RMS$_{1 ks}$ for the individual observations of \xmm, Chandra, \szk, see Tables~\ref{tab_suz_fit},\ref{tab_xmm_log},\ref{tab_ch_res}. In all panels the colour scheme corresponds to the superorbital phase ranges according to the legend.}
\label{fig_flux_phasa}
\end{figure}

\subsection{Flux variability  along the orbit}

Another interesting point is a  change in the flux variability scale along the orbit. Figure  \ref{fig_flux_phasa} top shows the fluxes detected by \sw in individual observations as a function of an orbital phase. Different colours correspond to different superorbital phases.

As it was proposed in \citet{chernyakova12} RXTE observations indicate the disruption of the Be-star disk at superorbital phase $\sim 0.6$. After this event the size of the disk is gradually increasing up to its next disruption. The similar behaviour is seen in \sw data, see middle panel of Fig.~\ref{fig_flux_phasa}.
The maximal flux variability can be naturally expected as the compact object intersects the clumpy outer regions of the disk. The shift of the RMS maximum from $\phi=0.55$ at superorbital phase $\Phi=0.8-1$  to $\phi=0.65$ at $\Phi=0.0-0.2$ and $\phi=0.75$ at $\Phi=0.2-0.4$ in this case can be interpreted as a gradual increase of the disk size at superorbital time scales.
At the same time green and blue points do not show clear orbital variation. Such a behaviour can be expected if the compact object spend the whole orbit inside the dense homogeneous regions of the near-to-maximum size disk. 
The evolution of the averaged over superorbital phase RMS (see orange points) also indicates that
close to periastron  the compact object is embedded into the smooth dense region of the Be star disk (most stable with respect to superorbital changes), while  closer to the apastron  the compact object moves in clumpy outskirts of the disk.

The lower panel shows the scatter of the flux in individual observations of \xmm/Chandra/Suzaku.The highest RMS values are located around the phase $\phi=0.6$ and are accompanied by a large RMS scattering. Similar to the \sw data, this can be interpreted as the highest clumps number and the strongest superorbital-scale variations of the disk close to apastron. Further evidence for the clumpy structure of the disk's outskirts can be seen in long \xmm(X6) and \szk(S1) observations performed around the phase $\phi=0.6$, where one sees the variability of the column density within a single observation, see Fig.~\ref{fig_con_split_suz},~\ref{fig_cont_xmm}.

\subsection{Variability on longer time scales}
\label{sect::var_long_timescales}

The long time scale variability of \lsi has been well studied at different energy ranges (e.g. \cite{chernyakova12,lsifermi13}). In addition to the previously known flux modulation with the orbital/superorbital phase, we notice here, that the short time scale variability pattern also varies along the orbit. This pattern can be quantified in terms of the power law index (slope) $\Gamma$ of the Fourier power spectrum density  (PSD) for each of the analysed observations. 

To obtain the PSD we processed each observation with the \textit{powspec} tool from the HEASoft software package. To account for the presence of the white noise in the measurements we fitted the resulting PSD with the broken power law:
\begin{equation}
  P(\nu) = \left\{ \begin{array}{lr}
                        P_0*(\nu/\nu_{br})^\alpha, & \nu < \nu_{br} \\
                        P_0*(\nu/\nu_{br})^\beta,  & \nu \geq \nu_{br}    \\
                      \end{array} \\
               \right.
\end{equation}

\begin{figure}
  \centering
  \includegraphics[width=\columnwidth]{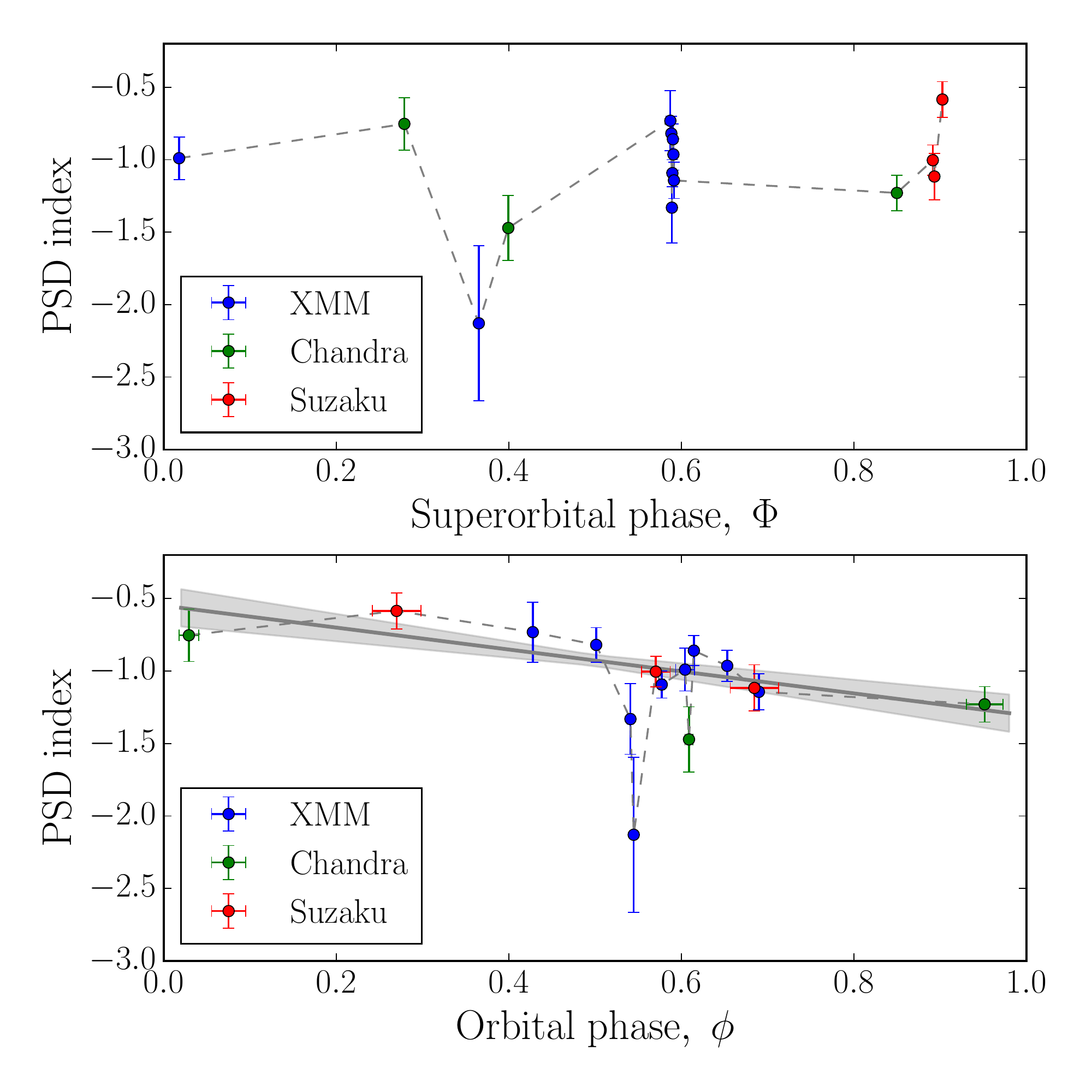}
  \caption{
    \textit{Upper panel:} evolution of the power  spectral density index~$\alpha$ of \lsi\ as a function of the superorbital phase.
    \textit{Lower panel}: the same as a function of the orbital phase.
    The dashed lines, joining the data points in both panels, are just to guide the eye. The continuous line and the grey shaded region represent the obtained best fit to the linear fit  together with its uncertainties.
  }\label{fig::flux_and_FFT_index}
\end{figure}

The obtained values of the lower-frequency index $\alpha$ are summarised in Fig.~\ref{fig::flux_and_FFT_index}, which shows their dependence on the orbital and superorbital phases.

For some of the data sets the obtained PSD was consistent with the white noise, so no estimate of $\alpha$ could have been obtained. The corresponding data points are thus not shown in Fig.~\ref{fig::flux_and_FFT_index}. Also not shown there is the outcome of the \sw observations -- due to their short durations a reliable estimation of their PSD was not possible. Grouping the \sw data in wide orbital phase bins we still find them consistent with a white noise; the resulting uncertainties are, though, too large to result in any inconsistency with the other measurements, displayed in Fig.~\ref{fig::flux_and_FFT_index}.

% --------------------------------
% *** PSD and flux variability ***
% --------------------------------

{This Figure indicates}  that the PSD slope follows a characteristic pattern and seems to be periodic with the orbital period of the binary. Over the orbit the slope gradually decreases from $\alpha \approx -0.6$ to $\alpha \approx -1.2$ and abruptly changes back to the initial value at phase $\phi \approx 0$ (see the lower panel of Fig.~\ref{fig::flux_and_FFT_index}). In order to estimate its significance we have performed the linear fit ($y=k(x-0.5)+b$) to the obtained data points, which resulted in $\chi^2 / d.o.f = 20.2/13$ for $k=-0.8 \pm 0.2$ and $b=-0.93 \pm 0.04$, compared to $\chi^2 / d.o.f = 36.0/14$ for the angular coefficient fixed at $k=0$. This corresponds to a~$\approx 4\sigma$ significance of the detected trend. The obtained best-fit line together with its uncertainties is shown in grey in the lower panel of Fig.~\ref{fig::flux_and_FFT_index}.

The hardest slope $\alpha \approx -0.6$ is found close to the periastron at phase $\phi = 0.275$. As the PSD slope characterises the relative power of the long and short time scale variability, this behaviour indicates the increasing fraction of the short time scale variations as  the compact object moves closer the Be star. The minimal contribution of the short variability {is found at $\phi \approx 0.55$, approximately 5~days before the apastron}.

It is reasonable to assume that the observed variability of the X-ray flux originates from inhomogeneities in the Be companion wind flow, resulting in a varying supply of material to the compact object. The measured PSD slope in this case simply reflects the short- (high frequencies) and long- (low frequencies) scale structure of the wind, encountered by  the compact object along its orbit. Given the frequency range $\nu \sim 10^{-4} - 10^{-2}$~Hz of our power density spectra and the orbital velocity of the compact object, these measurements are sensitive to wind inhomogeneities of $l = \mathrm{v}/\nu \sim 10^4-10^6$~km in size (assuming the average velocity of the compact object $\mathrm{v} \simeq 100$~km/s). 

This assumption provides a reasonable explanation of the observed behaviour of the PSD slope. As  the compact object moves towards the Be star, it encounters progressively more structured (on the scales $<10^6$~km) regions of the Be decretion wind flow. The first passage of  the compact object through the disk at $\phi \sim 0$ put it in a very inhomogeneous environment, as revealed by the sudden increase of the PSD slope to $\alpha \approx -0.7$. Further advance of the compact object brings it to an even more inhomogeneous region, potentially resulting from the disruption of the decretion of the disk by  the compact objects's gravity. Gradual growth of $\alpha$ (and the fraction of the short time scale variability) continues until the periastron, after which $\alpha$ starts to gradually decrease as the compact object moves away from the Be star to the less structured, distant regions of the orbit.

The absence of a clear correlation between the PSD $\alpha$ and the superorbital phase suggests that this picture does not qualitatively change with the gradual build-up of the Be disk. However, the sampling of the available observations is too sparse to make any firm conclusions on this matter.

%%%%%%%%%%%%%%%%%%%%%
\section{Discussion}
%%%%%%%%%%%%%%%%%%%%

\begin{figure}
\centering
\includegraphics[width=1.1\columnwidth]{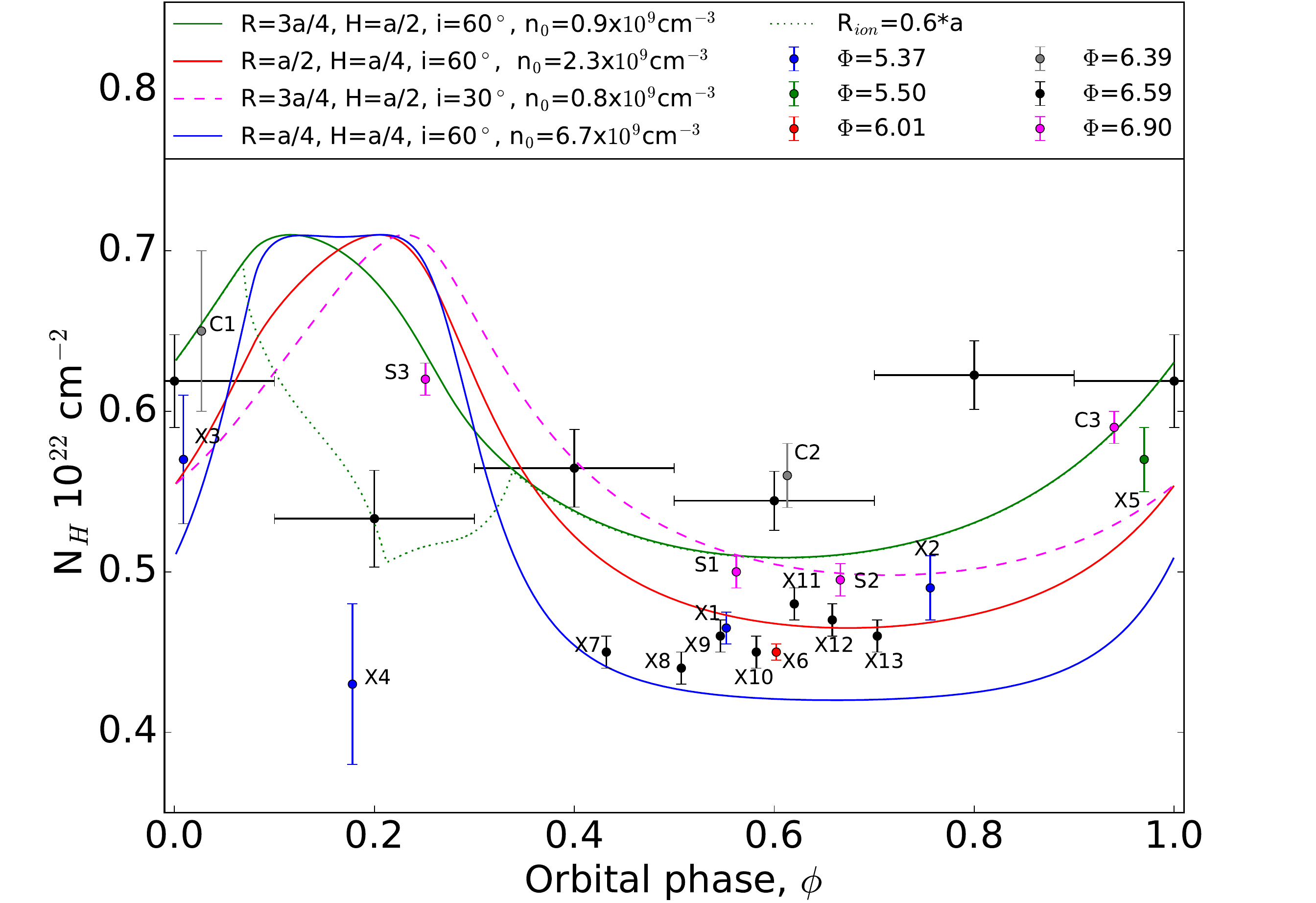}
\caption{$N_H$  orbital evolution in \lsi\ as observed by \textit{Suzaku (S1, S2 and S3), \textit{XMM-Newton} (X1 - X13)},  \textit{Chandra} (C1, C2, C3) and \textit{Swift} (black points with big horizontal error bars). Lines illustrates the $N_H$ behaviour predicted by the model described in  the text for different  sizes of the disk. Dotted green line illustrates the possible effect of the disk ionisation close to the star. The radius $R$ and the height scale of the disk $H$ are given in units of the major semiaxis $a$.} 
\label{fig_nh_phasa}
\end{figure}

{
The short and long time scale variability of \lsi\, reported here, provides important insights to the physical picture of the system, outlined in~\citet{chernyakova12}. The gradual build up and the subsequent destruction of the Be star disk constantly changes the conditions of medium, travelled through by the compact object. This all results in a different level of the disk fragmentation, which can be traced with the variability of the observed X-ray flux.

The presented observations demonstrate that both spectral and flux variability in \lsi\ happen already on the hour time scales. It is reasonable to assume, that this variability is caused not by the compact object itself, but rather by it probing the stellar wind regions of different structure (blob size and density) in the course of its movement around the star. 

This assumption is supported by the orbital and superorbital modulation of variability, illustrated in Fig.~\ref{fig_flux_phasa} in terms of the flux RMS. Indeed, in most of the MHD models of the Be star decretion disks their interior regions remain relatively homogeneous due to higher density, whereas the outskirts present highly structured, clumpy structure~\citep{Runacres02}. The orbital modulation of the flux RMS agrees with this picture, being lowest close to the epoch of periastron and gradually increasing towards the most distant part of the compact object's orbit at phase $\sim 0.7$. 

Further support to this picture is provided by the PDS variations, shown in Fig.~\ref{fig::flux_and_FFT_index}. Indeed, the gradual softening of the PDS index after the periastron passage suggests the small-scale blobs become rare -- as a result of their growth due to the internal pressure and the overall rarefaction of the disc.

In addition to this qualitative picture, the available data allow to obtain some quantitative information about the disk regions, interacting with the compact object. Though their structure can be assessed (at least to a certain extent) from the variability, the overall density of the disk can be obtained from the absorption signatures in the detected X-ray flux. 

In Figure~\ref{fig_nh_phasa} we summarise the measurements of the $N_H$ column density, stemming from the analysis described in Section 2. The value of $N_H$ is clearly non-constant along the orbit with a $\chi ^2=458.6$ for 18 degrees of freedom, suggesting the variability at the $19.6~\sigma$ level. Within the simplest model, the observed value of \nh at each orbital/superorbital phase is given by the integration of the smooth Be star disk density profile along the line of sight to the observer. The presence of clumps in the wind and/or the regions with highly or partially ionised hydrogen can, however, significantly modify the predictions for the observed \nh values.
}

We assume that the Be star disk consists of non-ionized hydrogen and  has an exponential density profile characteristic of isothermal atmosphere: $$n_D=n_0 \exp(-r/R-|z|/H)$$ In such a simple case one can expect to see a maximum of the column density when  the compact object crosses the periastron if an observer looks at the system in the direction perpendicular to the disk, or  when the  compact object passes the superior conjunction, if an observer is located in the orbital plane.  
In Figure~\ref{fig_nh_phasa} we show the \nh orbital evolution for different parameters of the disk (the contribution of the galactic \nh was set to $N_{Hgal}=0.42 \times 10^{22}$ cm$^{-2} $ for the best match to the data)\footnote{We would like to note, that this value is less than the total galactic \nh value in \lsi direction ($0.7-0.9 \cdot 10^{22}$~cm$^{-2}$) and is in agreement with the value that can be deduced from 3D hydrogen distribution maps used by GALPROP code \citep{galprop}.}. In this plot the solid lines correspond to the case of inclination $i=60^\circ$, and the dashed one to the case of $i=30^\circ$. As expected the orbital position of the maximum column density shifts toward the phase of periastron, $\phi=0.275$ with the decrease of the inclination of the observer.  %$\varphi=0.23$.
In all cases, the derived number density of the disk is $n_0 \sim 10^{-9} \mathrm{cm^{-3}}$ -- close that typically found for other Be stars~\citep{rivinius13}.

The small number of the available observations around the periastron does not allow to investigate the variations of $N_H$ in finer phase bins. Nevertheless we note that the nearby points S3 and X4 are marginally inconsistent. To explore this issue we have averaged all the \textit{Swift} data points presented in Table \ref{tab_swift_res_group} over the superorbital phases. The resulting profile, shown with black points in Fig.~\ref{fig_nh_phasa}, experiences the same drop of the column density at the orbital phase $\sim0.2$. A possible explanation of this behaviour could be an ionisation of the inner disk regions by the UV emission of Be star. To qualitatively test this possibility, we have assumed the fully ionised disk region of a size $R_{ion}$ does not contribute to the absorption of X-ray flux and repeated the modelling outlined above. The obtained $N_H$ profile, depicted with the green dotted line in Figure \ref{fig_nh_phasa}, shows the immediate drop of the observed \nh values around the periastron and its fast recovery afterwards, in concordance with the observed behaviour (the dotted and solid green lines in the figure correspond to the same configuration of the disk but with and without ionisation). Clearly, the available dataset is too limited to move beyond this qualitative picture; still it demonstrates that a dedicated observational campaign of the periastron passage may shed light on the ionisation state of the inner regions of the decretion disk.

\begin{figure}
\centering
\includegraphics[width=\columnwidth]{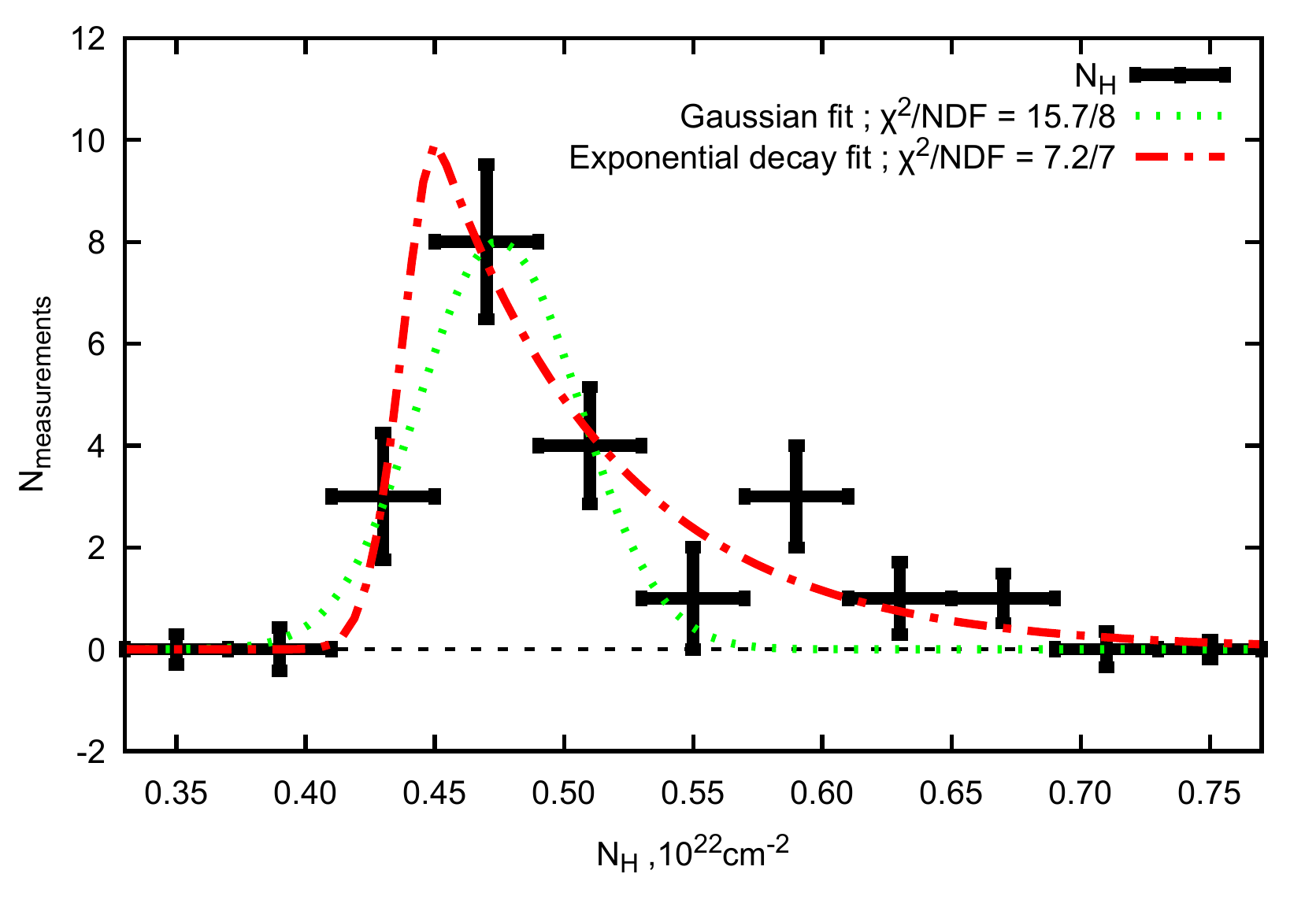}
\caption{The distribution of $N_H$ column densities for the \xmm, Chandra and Suzaku data shown in Fig.~\ref{fig_nh_phasa}. Green dashed and red dot-dashed curves correspond to the fit of the distribution with a single gaussian and ``gaussian rise, exponential decay'' profiles. The error bars for the distribution were estimated from $10^4$ random realisations of the original \nh dataset. In each random realization \nh values were selected to be gaussian-distributed random variables with the mean and dispersion given by the original distribution.}
\label{fig_nh_stat}
\end{figure}

At the same time, the observed distribution of $N_H$ can originate from the constant $N_H$ level modified by the presence of dense clumps that intersect the line of sight and effectively increase the $N_H$ on short time scales. From this point of view, the distribution of measured $N_H$ values\footnote{These data are summarized in Tables~\ref{tab_suz_fit},\ref{tab_xmm_res},\ref{tab_ch_res}. {Please note, that the observations S1a, S1b, and S2a, S2b were considered as individual data points}}, shown in Fig.~\ref{fig_nh_stat} may help to infer the properties of this clumpy medium.

The obtained distribution is best described with the ``gaussian rise, exponential decay'' profile, which may indicate the presence of another source of scatter in the data - in addition to the statistical one. This former can originate either from un-accounted for systematic errors in $N_H$ measurements or from a stochastic variation of the Be star disk density, with the dispersion exceeding the uncertainties of the presented data. In other words, the ``gaussian rise, exponential decay'' profile corresponds to the presence of the disk clumps with a certain distribution of densities/sizes.

In this case, the exponent decay scale $N_{H,dec}\approx 0.07\cdot 10^{22}$~cm$^{-2}$ would correspond to the characteristic column density of the clumps. {Assuming  mean density of the disk to be $n_0 \sim 10^9~\mathrm{cm^{-3}}$ \citep{rivinius13}}, this translates into the size of the clumps $s=N_H/n_0 \sim 7\cdot 10^{11}$~cm, comparable with the radius of the Be star. The obtained value is larger than the spatial scales $\sim 10^9-10^{11}$~cm, accessible from the variability detections, described in Sect.~\ref{sect::var_long_timescales}. This suggests that longer, $\sim 100$ ks continues observations are needed to fully characterise the structure of the disk. In addition to this, an accurate measurement of the disk density profile requires a much denser observational sampling along the orbit at different superorbital phases, clearly still accessible with the current generation of the X-ray instruments.
\newline
\newline
\noindent\textbf{Acknowledgements}

\noindent This work was partially supported by the EU COST Action (COST-STSM-MP1304-28864) ``NewCompStar''. The authors thank SFI/HEA Irish Centre for High-End Computing (ICHEC) for the provision of computational facilities and support. The work of IuB was partially supported  by the Stipendium of the President of Ukraine (2014-2016). DM was supported by the Carl-Zeiss Stiftung through the grant ``Hochsensitive Nachweistechnik zur Erforschung des unsichtbaren Universums'' to the Kepler Center f{\"u}r Astro- und Teilchenphysik at the University of T{\"u}bingen.

% Bibliography and bibfile
\def\aj{AJ}%
          % Astronomical Journal
\def\actaa{Acta Astron.}%
          % Acta Astronomica
\def\araa{ARA\&A}%
          % Annual Review of Astron and Astrophys
\def\apj{ApJ}%
          % Astrophysical Journal
\def\apjl{ApJ}%
          % Astrophysical Journal, Letters
\def\apjs{ApJS}%
          % Astrophysical Journal, Supplement
\def\ao{Appl.~Opt.}%
          % Applied Optics
\def\apss{Ap\&SS}%
          % Astrophysics and Space Science
\def\aap{A\&A}%
          % Astronomy and Astrophysics
\def\aapr{A\&A~Rev.}%
          % Astronomy and Astrophysics Reviews
\def\aaps{A\&AS}%
          % Astronomy and Astrophysics, Supplement
\def\azh{AZh}%
          % Astronomicheskii Zhurnal
\def\baas{BAAS}%
          % Bulletin of the AAS
\def\bac{Bull. astr. Inst. Czechosl.}%
          % Bulletin of the Astronomical Institutes of Czechoslovakia
\def\caa{Chinese Astron. Astrophys.}%
          % Chinese Astronomy and Astrophysics
\def\cjaa{Chinese J. Astron. Astrophys.}%
          % Chinese Journal of Astronomy and Astrophysics
\def\icarus{Icarus}%
          % Icarus
\def\jcap{J. Cosmology Astropart. Phys.}%
          % Journal of Cosmology and Astroparticle Physics
\def\jrasc{JRASC}%
          % Journal of the RAS of Canada
\def\mnras{MNRAS}%
          % Monthly Notices of the RAS
\def\memras{MmRAS}%
          % Memoirs of the RAS
\def\na{New A}%
          % New Astronomy
\def\nar{New A Rev.}%
          % New Astronomy Review
\def\pasa{PASA}%
          % Publications of the Astron. Soc. of Australia
\def\pra{Phys.~Rev.~A}%
          % Physical Review A: General Physics
\def\prb{Phys.~Rev.~B}%
          % Physical Review B: Solid State
\def\prc{Phys.~Rev.~C}%
          % Physical Review C
\def\prd{Phys.~Rev.~D}%
          % Physical Review D
\def\pre{Phys.~Rev.~E}%
          % Physical Review E
\def\prl{Phys.~Rev.~Lett.}%
          % Physical Review Letters
\def\pasp{PASP}%
          % Publications of the ASP
\def\pasj{PASJ}%
          % Publications of the ASJ
\def\qjras{QJRAS}%
          % Quarterly Journal of the RAS
\def\rmxaa{Rev. Mexicana Astron. Astrofis.}%
          % Revista Mexicana de Astronomia y Astrofisica
\def\skytel{S\&T}%
          % Sky and Telescope
\def\solphys{Sol.~Phys.}%
          % Solar Physics
\def\sovast{Soviet~Ast.}%
          % Soviet Astronomy
\def\ssr{Space~Sci.~Rev.}%
          % Space Science Reviews
\def\zap{ZAp}%
          % Zeitschrift fuer Astrophysik
\def\nat{Nature}%
          % Nature
\def\iaucirc{IAU~Circ.}%
          % IAU Cirulars
\def\aplett{Astrophys.~Lett.}%
          % Astrophysics Letters
\def\apspr{Astrophys.~Space~Phys.~Res.}%
          % Astrophysics Space Physics Research
\def\bain{Bull.~Astron.~Inst.~Netherlands}%
          % Bulletin Astronomical Institute of the Netherlands
\def\fcp{Fund.~Cosmic~Phys.}%
          % Fundamental Cosmic Physics
\def\gca{Geochim.~Cosmochim.~Acta}%
          % Geochimica Cosmochimica Acta
\def\grl{Geophys.~Res.~Lett.}%
          % Geophysics Research Letters
\def\jcp{J.~Chem.~Phys.}%
          % Journal of Chemical Physics
\def\jgr{J.~Geophys.~Res.}%
          % Journal of Geophysics Research
\def\jqsrt{J.~Quant.~Spec.~Radiat.~Transf.}%
          % Journal of Quantitiative Spectroscopy and Radiative Trasfer
\def\memsai{Mem.~Soc.~Astron.~Italiana}%
          % Mem. Societa Astronomica Italiana
\def\nphysa{Nucl.~Phys.~A}%
          % Nuclear Physics A
\def\physrep{Phys.~Rep.}%
          % Physics Reports
\def\physscr{Phys.~Scr}%
          % Physica Scripta
\def\planss{Planet.~Space~Sci.}%
          % Planetary Space Science
\def\procspie{Proc.~SPIE}%
          % Proceedings of the SPIE
\let\astap=\aap
\let\apjlett=\apjl
\let\apjsupp=\apjs
\let\applopt=\ao
\bibliographystyle{mn2e}
\bibliography{Pulsar_Catalog_ALL_Refs_new}

 \label{lastpage}

\end{document}